\documentclass{aa}

\usepackage{graphicx}
\usepackage{amsmath}
\usepackage{txfonts}
\usepackage[colorlinks=true,linkcolor=red,citecolor=blue]{hyperref}

\DeclareUnicodeCharacter{2212}{-}

\begin{document} 

\title{A deep 1.4 GHz survey of the J1030 equatorial field}
\subtitle{A new window on radio source populations across cosmic time}

   \author{Q. D'Amato \inst{1,2}\thanks{\email{qdamato@sissa.it}}
          \and
          I. Prandoni\inst{2}
          \and
          R. Gilli\inst{3}
          \and
          C. Vignali\inst{3,4}
          \and
          M. Massardi\inst{5,2,1}
          \and
          E. Liuzzo\inst{5,2}
          \and
          P. Jagannathan\inst{6}
          \and
          M. Brienza\inst{4,2}
          \and
          R. Paladino\inst{5,2}
          \and
          M. Mignoli\inst{3}
          \and
          S. Marchesi\inst{3,7}
          \and
          A. Peca\inst{8} 
          \and
          M. Chiaberge\inst{9} 
          \and
          G. Mazzolari\inst{4,3}    
          \and
        C. Norman\inst{9,10}
          }

   \institute{Scuola Internazionale Superiore di Studi Avanzati (SISSA), Via Bonomea 265, 34136, Trieste, Italy 
         \and
             Istituto di Radioastronomia (IRA), Via Piero Gobetti 101, 40129, Bologna, Italy
         \and
             Osservatorio di Astrofisica e Scienza dello Spazio di Bologna (OAS), Via P. Gobetti 93/3, 40129, Bologna, Italy
         \and
             Dipartimento di Fisica e Astronomia dell’Università degli Studi di Bologna, Via P. Gobetti 93/2, 40129, Bologna, Italy
         \and
             Italian Alma Regional Center (ARC), Via Piero Gobetti 101, 40129, Bologna, Italy
         \and
             National Institute of Radio Astronomy (NRAO), 801 Leroy Place Socorro, NM 87801, USA
         \and
             Department of Physics and Astronomy, Clemson University, Kinard Lab of Physics, Clemson, SC 29634, USA
         \and
             Department of Physics, University of Miami, Coral Gables, FL 33124, USA
         \and
            Space Telescope Science Institute, 3700 San Martin Drive, Baltimore, MD 21218, USA
         \and
            Department of Physics and Astronomy, Johns Hopkins University, Baltimore, MD 21218, USA
         }

\date{Received XXX; accepted XXX}

 
  \abstract
{We present deep L-Band observations of the equatorial field centered on the $z=6.3$ Sloan digital sky survey (SDSS) quasar (QSO). This field is rich of multiwavelength photometry and spectroscopy data, making it an ideal laboratory for galaxy evolution studies. Our observations reach a 1$\sigma$ sensitivity of ${\sim}$2.5 $\mu$Jy at the center of the field. 
We extracted a catalog of 1489 radio sources down to a flux density of ${\sim}$12.5 $\mu$Jy ($5\sigma$) over a field of view of ${\sim}$ $30'$ diameter. We derived the source counts accounting for catalog reliability and completeness, and compared them with others available in the literature. Our source counts are among the deepest available so far, and, overall, are consistent with recent counts' determinations and models. They show a slight excess at flux densities $\sim 50$ $\mu$Jy, possibly associated with the presence of known overdensities in the field. \\
We detected for the first time in the radio band the SDSS J1030+0524 QSO ($26 \pm 5 ~\mu$Jy, $8\sigma$ significance level). For this object, we derived an optical radio loudness $R_O =$ 0.62 $\pm$ 0.12, which makes it the most radio quiet among active galactic nuclei (AGN) discovered so far at $z\gtrsim6$ and detected at radio wavelengths.\\
We unveiled extended diffuse radio emission associated with the lobes of a bright Fanaroff-Riley type II (FRII) radio galaxy located close to the center of the J1030 field, which is likely to become the future brightest cluster galaxy of a protocluster at $z=1.7$. The lobes' complex morphology, coupled with the presence of X-ray diffuse emission detected around the FRII galaxy lobes, may point toward an interaction between the radio jets and the external medium.\\
We also investigated the relation between radio and X-ray luminosity for a sample of 243 X-ray-selected objects obtained from 500 ks Chandra observations of the same field, and spanning a wide redshift range ($0\lesssim z\lesssim 3$). Focused on sources with a spectroscopic redshift and classification, we found that sources hosted by early-type galaxies and AGN follow $\log(L_R)/\log(L_X)$ linear correlations with slopes of ${\sim}$0.6 and ${\sim}$0.8, respectively. This is interpreted as a likely signature of different efficiency in the accretion process. 
Finally, we found that most of these sources ($\gtrsim$87\%) show a radio-to-X-ray radio loudness $R_X \lesssim -3.5$, classifying these objects as radio quiet.
}
   

   \maketitle
%

\section{Introduction}
\label{sec:intro}
It is well known that the extra-galactic radio sky is dominated by two populations: star-forming galaxies (SFGs) and active galactic nuclei (AGN). \cite[e.g.,][]{miley_1980,condon_1992}. The radio loud (RL) AGN population dominates the bright radio sky ($S_{\rm 1.4 GHz} \gg$ 1 mJy), and, despite a steadily increasing contribution from SFGs, it remains the most relevant population down to flux densities of few hundreds of $\mu$Jy \cite[e.g.,][]{mignano_2008}; then SFGs take over (typically around $S_{\rm 1.4 GHz} {\sim}$200 $\mu$Jy; e.g., \citealt{simpson_2006,seymour_2008,smolcic_2008,novak_2017,prandoni_2018}). Recent studies have shown that at sub-mJy flux densities an increasing fraction of sources display signatures of radiatively efficient AGN activity, when observed at nonradio wavelengths \cite[e.g.,][]{bonzini_2013,padovani_2015}. These objects are often referred to as radio quiet (RQ) AGN since, generally, they do not show the usual RL AGN morphological features, such as strong relativistic radio jets and lobes\footnote{For this reason  \cite{padovani_2017} proposed to update the terms RL and RQ AGN into jetted and nonjetted AGN.}. It is interesting to note that RQ AGN overcome RL AGN at flux densities $\lesssim$ 100 $\mu$Jy (e.g., \citealt{bonzini_2013}). 
RQ AGN often have similar properties to SFGs, in terms of radio luminosity, radio luminosity function, host galaxy stellar mass, color, and morphology, suggesting that the radio emission in RQ AGN is to be ascribed to star formation \cite[e.g.,][]{bonzini_2013,bonzini_2015}. Several methods are used to reveal and quantify the presence of AGN-triggered radio emission in RQ AGN, such as for example very large baseline interferometry (VLBI) observations that allow one to reveal AGN radio cores and/or small-scale (parsecs and kiloparsecs) radio jets \citep{maini_2016,herrera_2016,herrera_2017}, and multiband analyses that allow one to identify the presence of a radio excess with respect to what is expected from pure star formation \cite[e.g.,][]{delvecchio_2017}. These studies have shown that 
RQ AGN showing radio cores or excess radio emission account for at least $\sim 30\%$ of the total RQ AGN population (e.g., \citealt{maini_2016,delvecchio_2017}). 
Deep radio surveys, sensitive to both RL and RQ AGN, as well as to SFGs,  represent an extremely powerful tool to investigate galaxy and AGN populations up to high redshift, as radio emission is insensitive to dust and gas obscuration \cite[see][and reference therein]{smolcic_2017b}. However, the physical properties and redshift distribution of radio sources cannot be determined without the availability of extensive multiband information. For example, Mid InfraRed (MIR) colors can be used to reliably separate SFGs from radiatively efficient AGN \citep{lacy_2004,stern_2005,donley_2012}. Spectral energy distribution (SED) fitting based on dense multiband photometry can provide reliable redshift estimates \citep[e.g.,][]{duncan_2018a,duncan_2018b}. Accurate redshifts can also be obtained by optical and Near IR (NIR) spectroscopy, when available, as well as line diagnostics for host galaxy classification \citep[e.g.,][]{baldwin_1981,veilleux_1987}. X-ray data can also be exploited to identify radiatively efficient AGN \citep{bonzini_2013,bonzini_2015}. In addition radio and X-ray investigations are key to test AGN-galaxy coevolution models \citep{lafranca_2010,bonchi_2013}. X-ray observations typically probe the innermost accretion processes, while the radio band can provide information on the jets. Radio jets are thought to inject a significant amount of energy and matter into the host galaxy and surrounding medium, likely affecting the subsequent evolution of the galaxy and of the central super massive black hole (SMBH; this process is called AGN feedback).  In the local Universe there is established evidence of interactions between the jets of radio galaxies (often found at the center of galaxy clusters) and the surrounding inter cluster medium (ICM), through the observation of so-called “X-ray cavities” (i.e., depression of the thermal X-ray emission of the hot gas in correspondence with the radio emission of the radio jets and lobes; e.g., \citealt{boehringer_1993,carilli_1994,birzan_2004}). In addition, the relation between the radio and X-ray luminosity has been extensively explored in the past for both RL and RQ AGN, leading to the discovery of a fundamental plane between the X-ray emission, the radio emission, and the SMBH mass \citep{merloni_2003,falcke_2004}. The X-ray and radio emission are generally found to follow a (logarithmic) linear correlation with slope distributed around ${\sim}$1 \citep{canosa_1999,brinkmann_2000,panessa_2007, fanbai_2016}. In particular, for low-luminosity AGN a standard value of 0.5--0.7 is found \citep{merloni_2003,dong_2021}, similarly to the typical slope of X-ray binaries \citep{narayan_1994}. This has been interpreted as a signature of radiatively inefficient accretion mechanisms (\citealt{gallo_2003,merloni_2003}), such asadvection-dominated accretion flows \citep{narayan_1994,heckman_2014}. Other works show a steeper relation (slope ${\sim}$1-4-1.6) for some X-ray binaries and for bright AGN, suggesting that in these objects a radiatively efficient accretion process is in place \citep{coriat_2011,dong_2014}, ascribed to a geometrically thin and optically thick accretion disk \citep{shakura_1973}. However AGN radio/X-ray studies are mainly based on local samples and clusters.
In order to shed light on accretion--feedback processes at high redshifts, it is necessary to exploit very deep radio and X-ray samples, with extended multiband ancillary data available, for a proper characterization of the source properties.\\


The field centered around the $z = 6.3$ Sloan digital sky survey (SDSS) quasar (QSO) J1030+0524 (RA = $10^h 30^m 27^s$, Dec = $+5^{\circ} 24' 55''$, \citeauthor{fan_2001} \citeyear{fan_2001}) represents an ideal laboratory to investigate AGN and galaxy coevolution up to high redshifts in different environments, as it is rich of galaxy overdensities. Over the past years, significant efforts have been made to collect deep multiwavelength photometric and spectroscopic data of this field\footnote{A comprehensive description of the available data can be found at \url{http://j1030-field.oas.inaf.it}.} (hereafter J1030). As a result the first evidence for an overdensity assembling around a powerful $z > 6$ AGN was found (\citeauthor{mignoli_2020} \citeyear{mignoli_2020}, see also \citeauthor{morselli_2014} \citeyear{morselli_2014} and \citeauthor{balmaverde_2017} \citeyear{balmaverde_2017}). 
 Particularly relevant is the fact that the J1030 field is covered by a ${\sim}$500 ks X-ray observation performed with ACIS-I instrument on board the \textit{Chandra} telescope, that is the fifth deepest extragalactic X-ray field to date \citep{nanni_2018}, from which a catalog of 256 extragalactic X-ray sources has been extracted \citep{nanni_2020}. 
By exploiting the exceptional multiwavelength coverage of the field and dedicated follow-up programs, \cite{marchesi_2021} derived the photometric (or, where possible, spectroscopic) redshift for 95\% (243/256) of the X-ray sources\footnote{We notice that 7 of the 13 sources with no redshift estimate are stars.}. Based on these data multiple galaxy structures were found in  the redshift range $z$ = 0.15--1.5. 

The field was observed for the first time in the radio band by \cite{petric_2003} with the very large array (VLA), achieving a resolution of ${\sim}$1.5'' and a root-mean-square (rms) of ${\sim}$15 $\mu$Jy at 1.4 GHz. Thanks to these observations, reanalyzed by \cite{nanni_2018}, the presence of a powerful, Compton-thick (X-ray derived column density $N_H{\sim} 1.5 \times 10^{24} \mathrm{cm^{-2}}$) radio galaxy, located ${\sim}$ 40'' southwest from the field center, was revealed. The radio source displays classical Fanaroff-Riley type II \cite[FRII;][]{fanaroff_1974} morphology, with an unresolved core and two bright lobes, where the western lobe (W-lobe) is $\gtrsim 6$ brighter than the eastern lobe (E-lobe). 

\cite{gilli_2019} measured the redshift of the FRII galaxy (z=1.7) and reported the discovery of an overdensity of SFGs assembling around the FRII galaxy, on the basis of spectroscopic data collected using the very large telescope (VLT) multi-unit spectroscopic explorer (MUSE) and Large binocular telescope (LBT) Utility Camera in the Infrared (LUCI). 
Exploiting Atacama large (sub-)millimeter array (ALMA) observations, \cite{damato_2020b} found that the FRII host galaxy is surrounded by a large molecular gas reservoir ($M_{H_2} \sim 2 \times 10^{11}~\mathrm{M_\odot}$), and found three additional gas-rich members of the structure ($M_{H_2} \sim 1.5-4.8 \times 10^{10}~\mathrm{M_\odot}$). \cite{damato_2020b} also show that the system will likely evolve into a local massive cluster ($M_{sys} \gtrsim 10^{14}~\mathrm{M_\odot}$) and that the FRII galaxy will evolve into the brightest cluster galaxy (BCG). Based on SED fitting \cite{damato_2021} found that the FRII host galaxy has a star formation rate (SFR) of ${\sim}$570 $\mathrm{M_{\odot}}$/yr and  a stellar mass of $M_{\ast} \sim 3.7 \times 10^{11}$ $\mathrm{M_{\odot}}$; the high corresponding specific SFR (sSFR) = $1.5 \pm 0.5$ $\mathrm{Gyr^{-1}}$ classifies this object as a starburst galaxy \citep{schreiber_2015} that will deplete its molecular gas reservoir in $\sim$ $3.5 \times 10^8$ yr. 

The deep \textit{Chandra} data \citep{nanni_2018} revealed several spots of diffuse X-ray emission around the FRII galaxy, where the most significant one overlaps with the E-lobe of the FRII galaxy. \cite{gilli_2019} proposed that this emission likely originates from an expanding bubble of gas in the ICM, that is shock-heated by the radio galaxy jet. Very interestingly, four of the MUSE protocluster members lie in an arc-like shape at the edge of the A spot. The sSFR measured in these galaxies is ${\sim}2-5\times$ that of the other protocluster members and that of typical main sequence galaxies of equal stellar mass and redshift. \cite{gilli_2019} suggest that the shock-heated expanding bubble may promote the star formation in these nearby galaxies. If confirmed, this would be the first evidence of positive AGN feedback on multiple galaxies at hundreds of kiloparsec scales \citep{gilli_2019}.

In this work we present new ${\sim}5\times$ deeper L-Band observations of the J1030 field, carried with the Karl G. Jansky VLA (JVLA). 
The observations are presented in Sec. \ref{sec:samp_sel}. The extracted source catalog and the source counts are presented in Sec. \ref{sec:cat} and  Sec. \ref{sec:num_counts}, respectively. To highlight the scientific potential of these data we then present some notable sources in Sec. \ref{sec:pec_sources}, and investigate the X-ray/radio luminosity relation for the X-ray-selected sample of \cite{marchesi_2021} in Sec. \ref{sec:X_rays}. We summarize our results in Sec. \ref{sec:concl}. Throughout the work we adopt a concordance $\Lambda$CDM cosmology with $\mathrm{H_0} = 70~\mathrm{km~s^{-1}~Mpc^{-1}}$, $\Omega_{\mathrm{M}} = 0.3$, and $\Omega_{\mathrm{\Lambda}} = 0.7$, in agreement with the \textit{Planck 2015} results \citep{PLANCK_2016}.

\section{Observations and data reduction}
\label{sec:samp_sel}

We observed the J1030 field at L-band (1 -- 2 GHz) with the JVLA in A configuration, in the period  May--June 2018 (Project ID: VLA/18A-440, PI: I. Prandoni), for a total on-source observing time of ${\sim}$30 hours (distributed in 11 observing blocks). The observed band is divided in 16 contiguous spectral windows (\textit{spws}); each \textit{spw} consists of 64 channels of 1 MHz width. The QSO 3C147 was used as flux and bandpass calibrator, while the QSO J1024-0052 served as phase calibrator\footnote{https://science.nrao.edu/facilities/vla/observing/callist}. We notice that the latter is a VLBI source, providing very accurate astrometry.

\begin{figure}[h!]
        \centering
\resizebox{\hsize}{!}
{\includegraphics[]{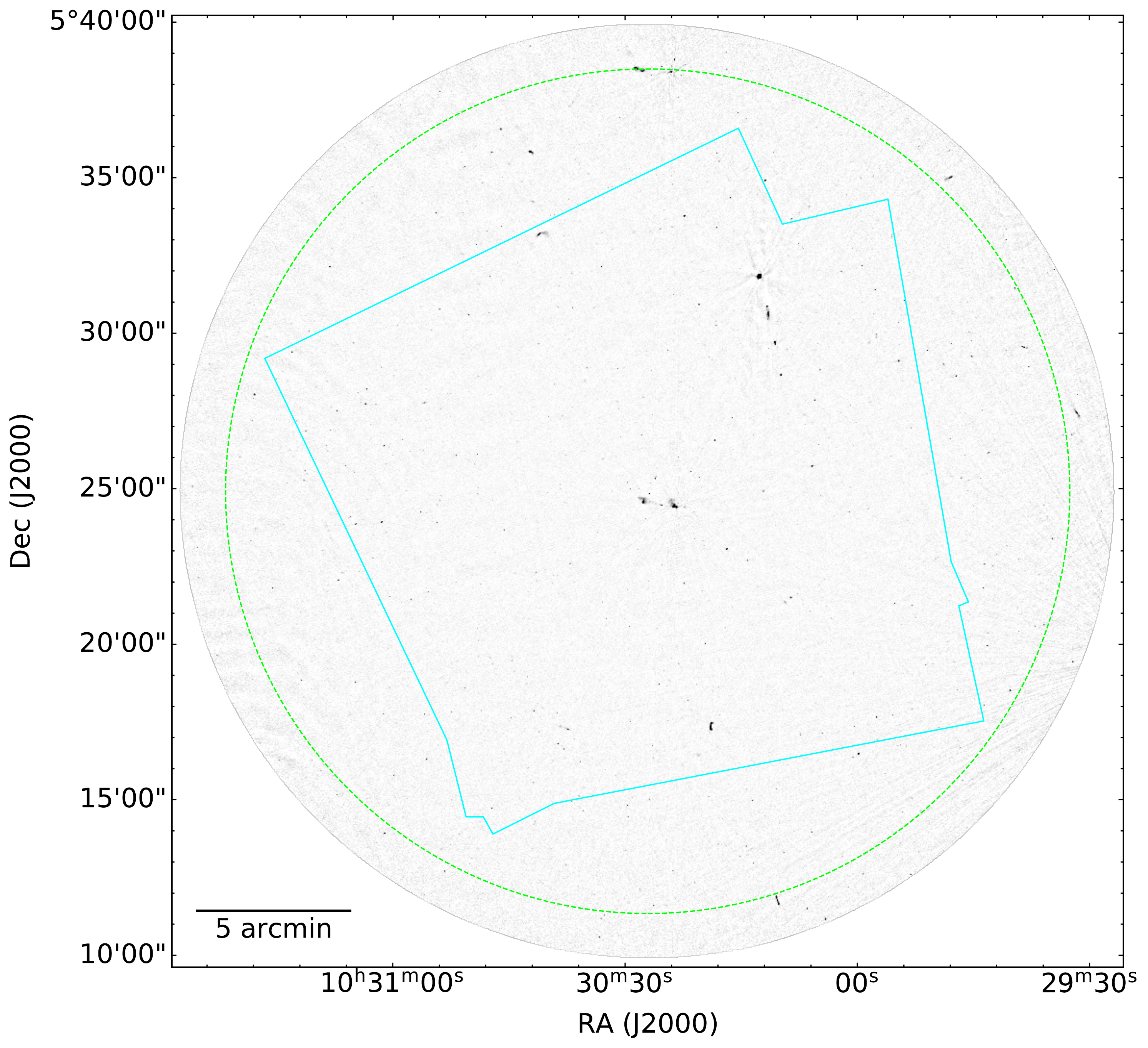}}

        \caption{JVLA image of the J1030 field. The  image has a diameter of $30^{\prime}$. The \textit{HPBW} of the observed field (${\sim}$27') is marked by the green dashed circle. The region covered by the \textit{Chandra} mosaic is marked by the solid cyan line.}
        
        \label{fig:field}
\end{figure}

The calibration and flagging  of the datasets was initially performed through the calibration pipeline of the common astronomy software applications (CASA) package \cite[version 5.1.2-4;][]{mcmullin_2007}. Unfortunately all the datasets were found to be strongly affected (>50\%) by radio frequency interference (RFI), as expected for the JVLA L-band\footnote{https://science.nrao.edu/facilities/vla/docs/manuals/obsguide/rfi}. This resulted in the failure of the calibration pipeline and required careful manual flagging and calibration of each dataset. Then, we restarted from the raw data and applied the Hanning--smoothing and online flags, following the pipeline procedure. Subsequently we calibrated the two calibrators in order to perform a preliminary RFI flagging both in time and frequency using the {\sc{flagdata}} task of CASA. We recursively repeated this step two times to refine the flags after each calibration. Finally, we inspected the data to perform small additional manual flags when required. We note that this procedure is especially important for the phase calibrator, since it traces the RFI's variation across the observing time of the target. Exploiting the high signal-to-noise ratio (S/N) of the calibrator, we were able to identify the most relevant RFI contamination across the frequency band and observing time.  We found that the \textit{spw} 8 was strongly affected by RFI and decided to flag it entirely. Then, we performed standard gain and bandpass calibration and applied the tables to the target. The calibrator flags are included in the final gain table and then are automatically applied to the target. 
We corrected the entire dataset for the leakage and instrumental cross-hand delays, by exploiting the unpolarized calibrator 3C147; this procedure led to a significant amount (${\sim}$10\%) of the total intensity flux retrieved from the leakage losses. 
We found that four out of the eleven observations were still strongly contaminated by the interference, and we decided to exclude them from the analysis. Then, we combined the remaining seven observations into the final dataset.
Finally, the dataset was self-calibrated by exploiting the luminous ($S_{\mathrm{1.4~GHz}} \sim 200$ mJy) National radio astronomy observatory (NRAO) VLA Sky Survey \cite[NVSS,][]{condon_1998} QSO J102921+051938, located in the J1030 field at ${\sim}$17 arcmin southwest from the field center.
We performed the imaging using the {\sc{tclean}} task of CASA with the {\sc{awproject}} gridder, as required for wide fields. The robust weighting was  set to 0.5, corresponding to a restoring beam with a major (minor) axis of 1.48 (1.15) arcsec. This choice corresponds to an optimal trade--off between resolution and sensitivity. We present the final image in Fig. \ref{fig:field}. The pixel scale is set to 0.3 arcsec. The half primary beam width (\textit{HPBW}) of the observed field is ${\sim}$27 arcmin, and is shown by the dashed green circle in Fig. \ref{fig:field}. 

\subsection{Time and bandwidth smearing}
\label{sec:smearing}
Radio images are in general affected by smearing, a systematic  effect that causes the distortion of the source morphology, by distributing the source emission on a wider area. Then, while the total flux is conserved, the peak emission of the source is reduced. Smearing increases with distance from the phase center of the image, and depends on the image spatial resolution, the observing frequency, and the time and frequency resolution of the observations. In particular, at a given distance from the image center, time and frequency resolutions determine distortions along the tangential and radial directions, respectively \cite[][Eq. 18-29 and Eq. 18-43]{bridle_1999}. Considering the integration time (2 seconds) and the spectral resolution (1 MHz) of our observations, we found that at 15 arcmin from the center the tangential and radial smearing are ${\sim}$0.3\% and ${\sim}$10\%, respectively. A radial smearing of 10\% is significant, but still acceptable. Considering that the image sensitivity degrades very steeply beyond the HPBW (green circle in Fig. \ref{fig:field}), we decided to limit our analysis to the inner 15 arcmin radius region (or inner $30^{\prime}$ diameter). This region fully encompasses the Chandra mosaic (solid cyan line in Fig. \ref{fig:field}). 

\section{Source detection and catalog extraction}
\label{sec:cat}
\begin{figure}[t]
        \centering
\resizebox{\hsize}{!}
{\includegraphics[]{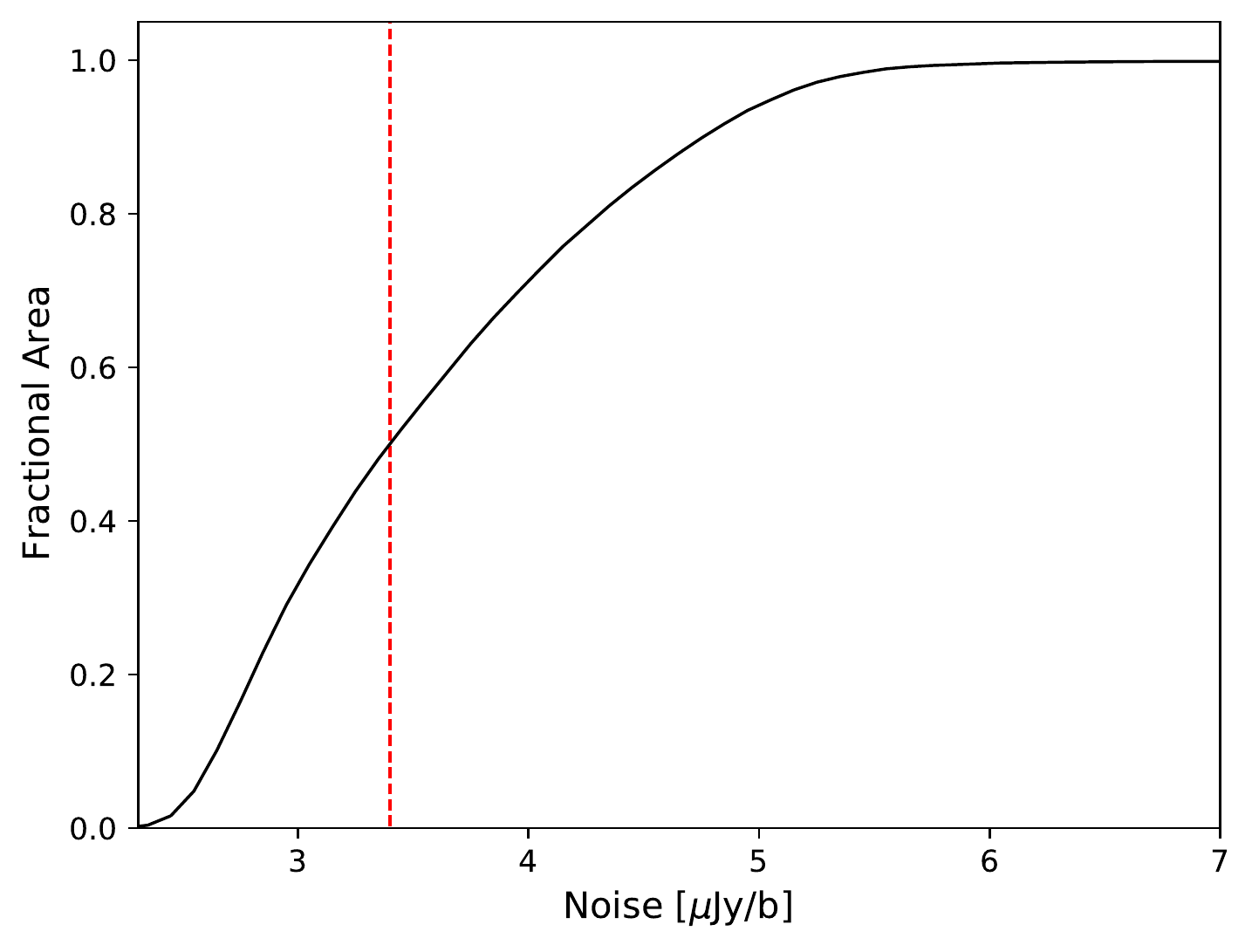}}

        \caption{Visibility area of the inner 15 arcmin radius of the image, corresponding to the region of the catalog extraction (black solid line). The noise level that encompasses the 50\% of the analyzed area (3.4 $\mu$Jy/b) is marked by the red vertical dashed line.}
        
        \label{fig:visarea}
\end{figure}

\begin{figure*}[t]
        \centering
{\includegraphics[scale=0.9]{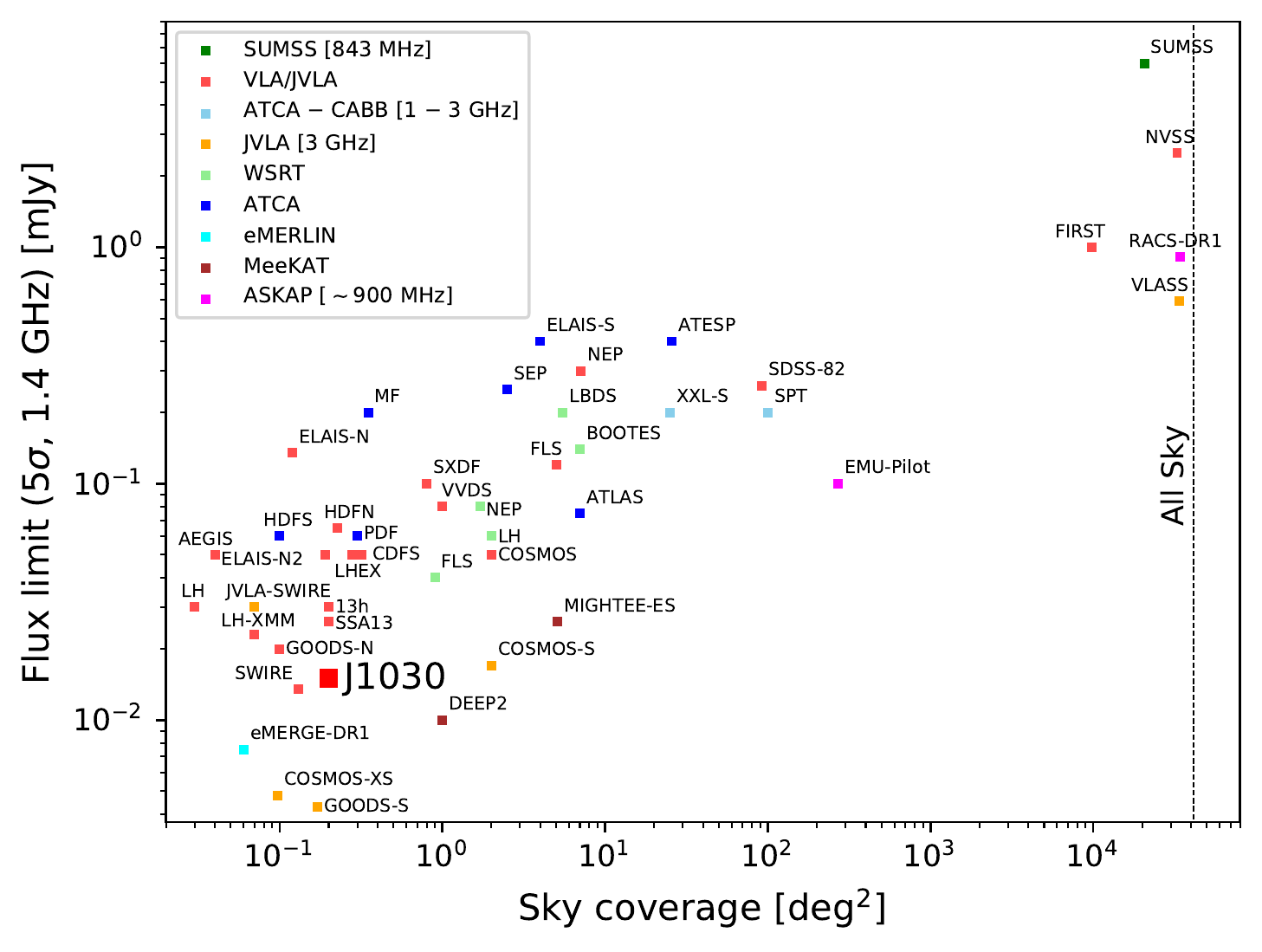}}
        \caption{Sky coverage versus 5${\sigma}$ sensitivity for a collection of available ${\sim}$GHz radio surveys to date, color-coded by the instrument as in the legend. The sensitivities refer to 1.4 GHz; the surveys at other frequencies have been rescaled to 1.4 GHz assuming a radio spectral index of $-$0.7. The large red square indicates the J1030 field presented in this work. The reported surveys are as follows: LBDS \citep{windhorst_1984}, NEP \cite[VLA,][]{kollgaard_1994}, FIRST \citep{white_1997}, MF \citep{gruppioni_1997}, NVSS \citep{condon_1998}, PDF \citep{hopkins_1998}, ELAIS--S \citep{gruppioni_1999}, ELAIS--N \citep{ciliegi_1999}, SUMSS \citep{bock_1999}, ATESP \citep{prandoni_2000}, BOOTES \citep{devries_2002}, VVDS \citep{bondi_2003}, FLS \cite[VLA,][]{condon_2003}, FLS \cite[WSRT,][]{morganti_2004}, 13h \citep{seymour_2004}, HDFN \citep{huynh_2005}, LHEX \citep{oyabu_2005}, SXDF \citep{simpson_2006}, ELAIS--N2, LH--XMM and HDFN \citep{biggs_2006}, SSA13 \citep{fomalont_2006}, ATLAS \citep{norris_2006}, COSMOS \citep{schinnerer_2007}, AEGIS \citep{ivison_2007}, SWIRE \citep{owen_2008}, LH \cite[VLA,][]{ibar_2009}, NEP \cite[WSRT,][]{white_2010}, GOODS--N \citep{morrison_2010}, SDSS--82 \citep{hodge_2011}, JVLA--SWIRE \citep{condon_2012}, SEP \citep{white_2012}, CDFS \citep{miller_2013}, COSMOS--S \citep{smolcic_2017}, LH \cite[WSRT,][]{prandoni_2018}, RACS--DR1 \citep{mcconnell_2020}, MIGHTEE--ES \citep{heywood_2021}, DEEP2 \citep{matthews_2021}, EMU--Pilot \citep{norris_2021}, COSMOS-XS \citep{van_der_vlugt_2021}, GOODS-S \citep{lyu+2022}, VLASS \cite[ongoing,][]{lacy_2016}, eMERGE--DR1 \cite[ongoing,][]{muxlow_2020}, SPT (ongoing, PI: N. Tothill), XXL--S (ongoing as part of SPT, PI: V. Smol{\v{c}}i{\'c}).
        }
        
        \label{fig:fields_comp}
\end{figure*}

We extracted the source catalog by using the PYthon Blob Detector and Source Finder \cite[PyBDSF;][]{mohan_2015}. We restricted our extraction to a region of $30^{\prime}$ diameter to keep bandwidth smearing under control, as described in Sec. \ref{sec:smearing}. 
The detection threshold was initially set to 4.5${\sigma}$, where ${\sigma}$ is defined as the local noise at the source position. This is a conservative value used by PyBDSF to search for potential sources, based on the flux distribution on the image. Once the source fitting is performed, the source S/N was recalculated based on the fitted peak flux, and a final detection threshold of 5${\sigma}$ was set based on the reliability analysis presented in Sec. \ref{sec:rel}. 
To characterize the image noise, we used a sliding box with a side of 45 pixels and a step of 15 pixels. The noise is calculated as the rms of pixel values within the box. This choice was found to fairly reproduce the noise variations measured across the image. In the regions around very bright sources (S/N $\gtrsim$ 150), noise variations are  stronger due to residual phase errors still present in the image. Hence the box side was reduced to 30 pixels with a step of 10 pixels. The visibility area of the region where the source catalog was extracted is shown in Fig. \ref{fig:visarea}. We note that the lowest noise value, measured at the center of the image, is ${\sim}$2.5 $\mu$Jy/b, that ${\sim}$50\% of the analyzed area has a noise level $\lesssim$ 3.4 $\mu$Jy/b (vertical red dashed line), and that the full area  is characterized by noise values below 5.5 $\mu$Jy/b. This makes the J1030 field one of the deepest radio survey to date; in Fig. \ref{fig:fields_comp} we compare the $5\sigma$ sensitivity and area coverage of the J1030 radio survey with the ones of other deep GHz surveys available to date.

\begin{table*}
\caption{\label{tab:cat} First ten entries of the radio source catalog.
}
\centering
\resizebox{\hsize}{!}{
\begin{tabular}{ccccccccccccc}
\hline \hline
RID & RA & Dec & $S_T$ & $S_P$ & Maj & Min & PA & $\mathrm{Maj_{DC}}$ & $\mathrm{Min_{DC}}$ & $\mathrm{PA_{DC}}$ & rms & Flag\\
   & [deg] & [deg] & [$\mu$Jy] & [$\mu$Jy/beam] & [arcsec] & [arcsec] & [deg] & [arcsec] & [arcsec] & [deg] & [$\mu$Jy/beam] & \\
\hline
1 & 157.86126 $\pm$ 4E-05 & 5.41278 $\pm$ 3E-05 & 51 $\pm$ 13 & 30 $\pm$ 5 & 2.2 $\pm$ 0.4 & 1.3 $\pm$ 0.2 & 125 $\pm$ 18 & 1.7 $\pm$ 0.4 & 0.5 $\pm$ 0.2 & 118 $\pm$ 18 & 4.7 & S \\ 
2 & 157.85841 $\pm$ 3E-05 & 5.38999 $\pm$ 3E-05 & 46 $\pm$ 11 & 36 $\pm$ 5 & 1.7 $\pm$ 0.3 & 1.3 $\pm$ 0.2 & 123 $\pm$ 24 & 1.0 $\pm$ 0.3 & 0.1 $\pm$ 0.2 & 102 $\pm$ 24 & 5.2 & S \\ 
3 & 157.85790 $\pm$ 3E-05 & 5.39102 $\pm$ 2E-05 & 105 $\pm$ 16 & 55 $\pm$ 6 & 2.3 $\pm$ 0.3 & 1.4 $\pm$ 0.1 & 117 $\pm$ 12 & 1.8 $\pm$ 0.3 & 0.7 $\pm$ 0.1 & 109 $\pm$ 12 & 5.5 & S \\ 
4 & 157.85756 $\pm$ 1E-05 & 5.41787 $\pm$ 9E-06 & 150 $\pm$ 12 & 86 $\pm$ 5 & 2.1 $\pm$ 0.1 & 1.4 $\pm$ 0.1 & 112 $\pm$ 7 & 1.6 $\pm$ 0.1 & 0.6 $\pm$ 0.1 & 100 $\pm$ 7 & 4.3 & S \\ 
5 & 157.85592 $\pm$ 2E-05 & 5.39921 $\pm$ 2E-05 & 41 $\pm$ 9 & 38 $\pm$ 5 & 1.6 $\pm$ 0.2 & 1.2 $\pm$ 0.1 & 136 $\pm$ 20 & -- & -- & -- & 4.9 & S \\ 
6 & 157.85487 $\pm$ 2E-05 & 5.38497 $\pm$ 1E-05 & 92 $\pm$ 11 & 68 $\pm$ 5 & 1.8 $\pm$ 0.2 & 1.3 $\pm$ 0.1 & 121 $\pm$ 11 & 1.2 $\pm$ 0.2 & 0.2 $\pm$ 0.1 & 104 $\pm$ 11 & 4.9 & S \\ 
7 & 157.85454 $\pm$ 4E-05 & 5.36900 $\pm$ 3E-05 & 110 $\pm$ 19 & 40 $\pm$ 5 & 2.5 $\pm$ 0.3 & 1.9 $\pm$ 0.2 & 111 $\pm$ 25 & 2.1 $\pm$ 0.3 & 1.4 $\pm$ 0.2 & 100 $\pm$ 25 & 4.9 & S \\ 
8 & 157.85355 $\pm$ 2E-05 & 5.39502 $\pm$ 1E-05 & 106 $\pm$ 14 & 64 $\pm$ 6 & 2.0 $\pm$ 0.2 & 1.4 $\pm$ 0.1 & 97 $\pm$ 14 & 1.5 $\pm$ 0.2 & 0.4 $\pm$ 0.1 & 85 $\pm$ 14 & 5.2 & S \\ 
9 & 157.85350 $\pm$ 3E-05 & 5.44232 $\pm$ 3E-05 & 48 $\pm$ 12 & 35 $\pm$ 6 & 1.6 $\pm$ 0.3 & 1.5 $\pm$ 0.2 & 86 $\pm$ 104 & -- & -- & -- & 5.2 & S \\ 
10 & 157.84849 $\pm$ 1E-05 & 5.48335 $\pm$ 1E-05 & 108 $\pm$ 11 & 70 $\pm$ 5 & 1.7 $\pm$ 0.1 & 1.6 $\pm$ 0.1 & 123 $\pm$ 37 & 1.1 $\pm$ 0.1 & 0.7 $\pm$ 0.1 & 73 $\pm$ 37 & 4.5 & S \\

\hline    	  
\end{tabular}
}
\tablefoot{All the columns of the full catalog are described in Sec \ref{sec:rel}. A complete version of the catalog will be available at the CDS.
}
\end{table*}

\subsection{Catalog reliability}
\label{sec:rel}
\begin{figure}[t]
        \centering
\resizebox{\hsize}{!}
{\includegraphics[]{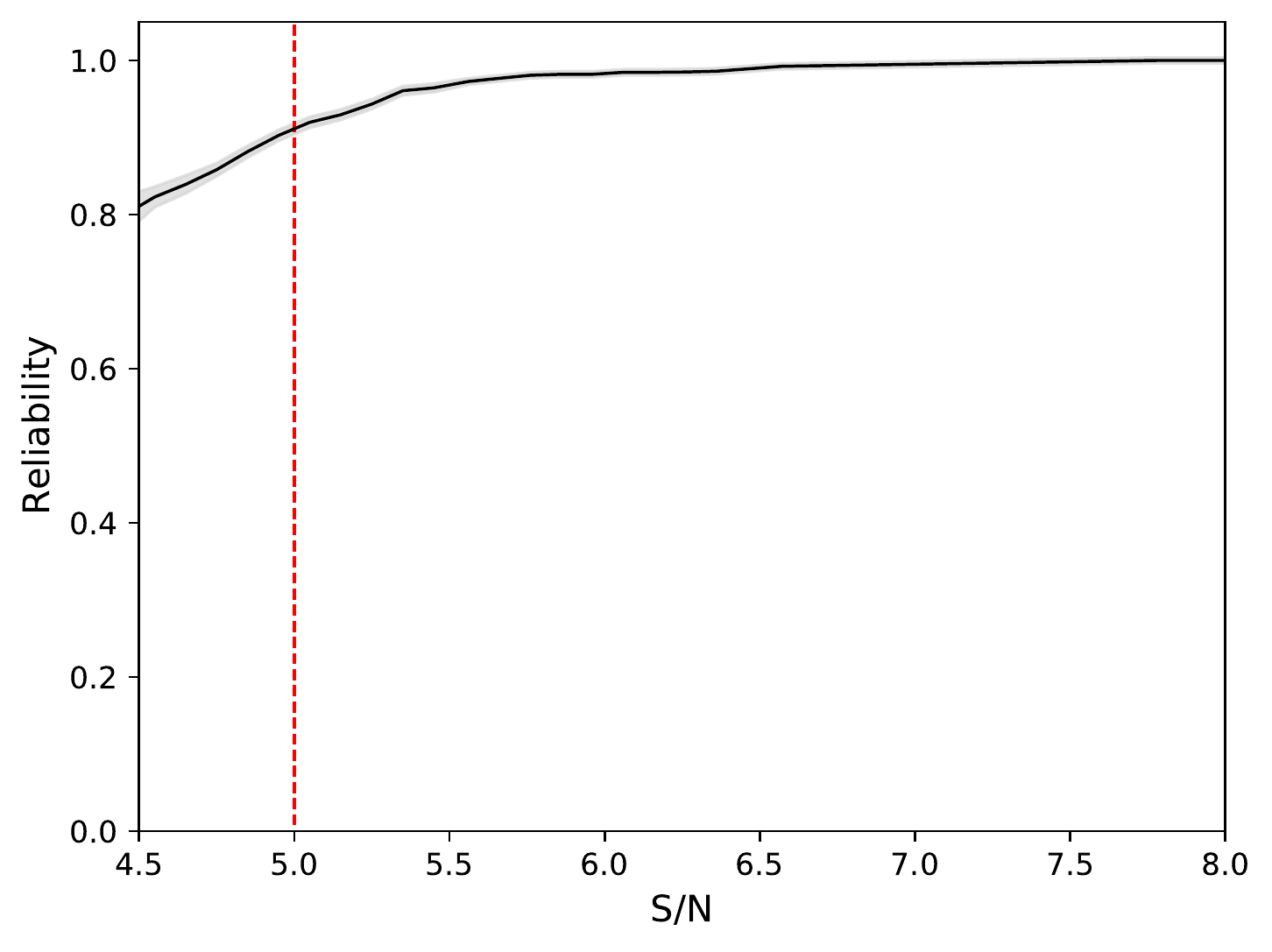}}

        \caption{Reliability function of the catalog (black solid line) with the 1-$\sigma$ uncertainty (gray shades) as a function of the S/N. The red vertical dashed line marks the detection threshold (S/N = 5), corresponding to a reliability value of 93\%.}
        
        \label{fig:relab}
\end{figure}

We assessed the reliability of our catalog as a function of the S/N. In order to do so, we ran PyBDSF on the negative image (i.e., an image in which each pixel has the same value, but inverted sign) using the same parameters adopted for the catalog extraction. This procedure provides us with the incidence and S/N distribution of spurious detections. We found a total of 14 negative detections with S/N $\geq$ 5 in the inner 15 arcmin radius region: they present a homogeneous spatial distribution, meaning that they are not clustered around bright sources, where the noise may significantly deviate from a random Gaussian distribution due to residual artifacts. We defined the false detection rate (FDR) as the ratio between the negative and positive detections, and the reliability as R = 1-$\mathrm{FDR|_{S/N}}$, where $\mathrm{FDR|_{S/N}}$ is the integrated FDR at a given S/N (i.e., FDR($>$S/N)). The reliability as a function of S/N is reported in Fig. \ref{fig:relab}. The errors are propagated from the Poisson errors of the negative detections. We found that the reliability is ${\sim}$93\% for S/N = 5, which is set as the final threshold for the catalog. Above this threshold we found a total of 1489 sources. The sources classified as multicomponent sources (i.e., the sources that are fitted by multiple Gaussian models) are 50, and are labeled by an ``M'' flag in the catalog. We inspected these sources and found that peak and total flux densities are consistent with manual measurements, while the Full width Half Maximum (FWHM) of Gaussian fits do not always provide reliable sizes for these multicomponent sources. 
In addition, we found 13 extended sources with a complex morphology, that clearly can not be described by the Gaussian fitting performed by PyBDSF. Many of these sources were also split into more entries of the catalog; we removed such entries and substituted them with a single entry for which we manually measured the flux density and size. These  sources are flagged as ``E'' in the catalog. The remaining 1426 sources are well-fitted by a single Gaussian component and are flagged as ``S'' in the catalog. We present the first ten entries of the source catalog in Table \ref{tab:cat}. A complete version of this table will be available at the Centre de Données astronomiques de Strasbourg (CDS). The details of the columns are described below:
\begin{list}{-}{}
\item Column 1: Radio source identification number (RID).
\item Columns 2 and 3: Right ascension (RA) and its error in degrees.
\item Columns 4 and 5: Declination (Dec) and its error in degrees.
\item Columns 6 and 7: Total flux density ($S_T$) and its error in $\mu$Jy.
\item Columns 8 and 9: Peak flux density ($S_P$) and its error in $\mu$Jy/beam.
\item Columns 10 and 11: Fitted major axis (Maj) and its error in arcsec.
\item Columns 12 and 13: Fitted minor axis (Min) and its error in arcsec.
\item Columns 14 and 15: Fitted position angle (PA) and its error in degrees.
\item Columns 16 and 17: Deconvolved major axis ($\mathrm{Maj_{DC}}$) and its error in arcsec.
\item Columns 18 and 19: Deconvolved minor axis ($\mathrm{Min_{DC}}$) and its error in arcsec.
\item Columns 20 and 21: Deconvolved position angle ($\mathrm{PA_{DC}}$) and its error in degrees.
\item Column 22: Local noise in $\mu$Jy/beam.
\item Column 23: Source model flag.
\end{list}

\subsection{Catalog completeness}
\label{sec:SNomp}
\begin{figure}[t]
        \centering
\resizebox{\hsize}{!}
{\includegraphics[]{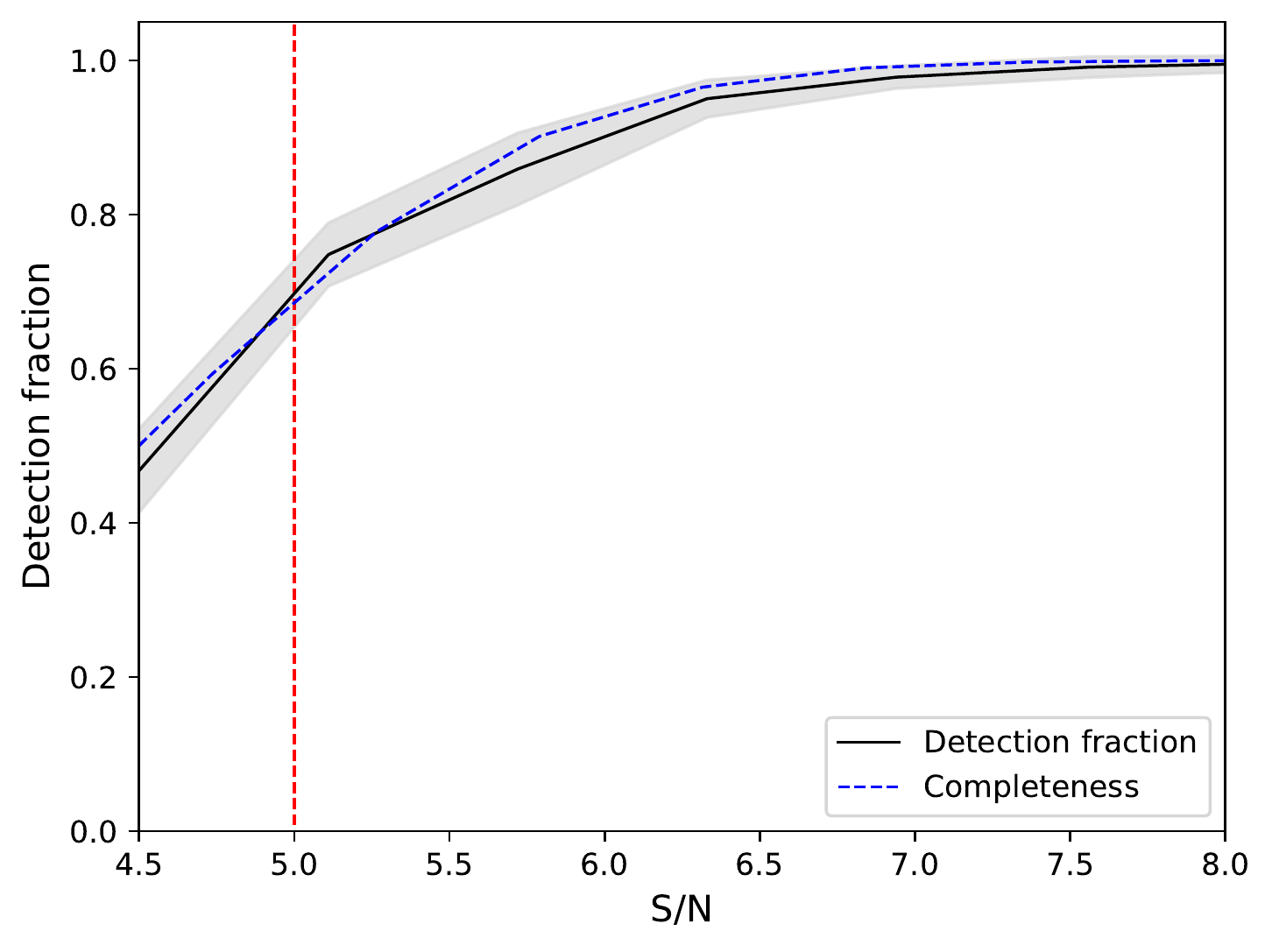}}

        \caption{Mean detection fraction of the point-like simulated catalogs (black solid line) with  1-$\sigma$ uncertainty (gray shades) as a function of S/N. The red vertical dashed line marks the detection threshold (S/N = 5). The blue dashed line indicates the S/N completeness function (see Eq.~\ref{eq:csnr}).}
        
        \label{fig:df}
\end{figure}

The presence of background noise introduces errors in the measurement of source fluxes. This causes source catalogs to be affected by incompleteness at low S/N. If the noise (and hence flux error) follows a Gaussian distribution, the S/N incompleteness can be analytically evaluated and corrected by exploiting the Gaussian error function (see, e.g., \citealt{mandal_2021}). In order to verify whether the image noise follows a Gaussian distribution, we generated ten simulated catalogs of 1000 point-like sources each, and used the {\sc{addcomponent}} tool of CASA to inject them in the image at random positions, over the same area where the source catalog was extracted (30 arcmin diameter). The sources are however required to be separated from each other by at least 4 arcsec, to avoid overlap. We checked that the visibility area is not significantly altered after the injection of these additional sources, meaning that the noise distribution is the same as in the original map. 
The fluxes of the simulated sources are drawn from an uniform distribution, in the range from $4.5\times$ the minimum  noise  to $10\times$ the maximum noise in the image. The uniform distribution has been chosen to have enough statistics in each bin of S/N. 
For each simulated catalog, we ran PyBDSF with the same parameters used for the extraction of the real catalog. The detected sources are defined by matching the PyBDSF detections with the input catalog sources, using a maximum separation of 0.5 arcsec between the injected source position and the PyBDSF position. 
Then, we calculated the detection fraction (DF) as a function of S/N as the ratio between the detected sources and the number of  simulated sources in each S/N bin (which is 0.5 wide). Finally, we calculated the overall DF and its error as the mean and standard deviation of the DFs of the ten simulations, respectively. We report the DF as a function of S/N in Fig. \ref{fig:df} (black solid line). We note that, since the DF has been calculated in bins of S/N, we do not have to correct it for the visibility area.
In the case of a pure Gaussian noise distribution, the probability of a single measurement to fall outside of a given range $\pm x$ is given by the complementary Gaussian error function $\mathrm{erfc}(x) = 1 - \mathrm{erf}(x)$, where $\mathrm{erf}(x)$ is the Gaussian error function. If we set $x = (N_{\mathrm{th}} - S/N)/\sqrt{2}$, where $N_{\mathrm{th}}$ = 4.5 is the detection threshold used for the catalog extraction, we can define the point source completeness as a function of S/N as:
\begin{equation}\label{eq:csnr}
C_\mathrm{S/N} = 0.5 \mathrm{erfc}(x)
\end{equation}
that is the probability of a measurement whose intrinsic S/N is below (above) the detection threshold to fall above (below) it. We report the completeness function (blue dashed line) in Fig. \ref{fig:df}, and compare it to the mean DF. This comparison shows that the two curves  are in agreement with each other (within 1{$\sigma$}; see gray shades), and thus that the point source completeness can be fairly predicted by assuming a Gaussian noise distribution. We note that in case of a Gaussian noise distribution, we can also analytically correct the catalog for the so-called resolution bias, that is the additional incompleteness affecting extended sources (Sec. \ref{sec:res_bias}).

\subsection{Effective frequency, source flux and position accuracy}
\label{sec:check_flux}
Continuum source flux densities derived from radio surveys are generally referred to the nominal central frequency of the observing band, that in our case is 1.5 GHz. However, the flux density is a quantity that can vary with frequency, and for a given source spectral index there is an effective frequency at which the image flux density equals the actual flux density of the source. In addition, in case of large bandwidths, the image flux density has to be weighted for the different primary beam response as a function of the frequency, so that a weighted effective frequency of the observation can be derived,  as the frequency at which the weighted image flux density is equal to the actual source flux density (see Eq. 15-17 of \citealt{condon_2015}; see also \citealt{heywood_2022}). We calculated that the weighted effective frequency of our observations is 1.34 GHz. 

We checked the flux scale of the catalogd sources by comparing the total flux of our sources against the values reported by the VLA Faint Images of the Radio Sky at Twenty-centimeters (FIRST) survey at 1.4 GHz \citep{becker_1995}. In order to increase the statistics we extended the PyBDSF source extraction to a larger area of 1 degree  diameter, under the assumption that total source fluxes are not affected by smearing. We used a matching radius of 2 arcsec, and restricted the match to the FIRST point-like sources ($|S_T/S_P| \leq 1.1$). 
We found that the median of the ratio between the total fluxes of the JVLA and FIRST catalogs is 1.02, fully consistent with the 1.03 ratio expected when comparing 1.34 and 1.4 GHz flux densities, assuming a spectral index $\alpha=-0.7$ (where $S_\nu \propto \nu^\alpha$).

We checked the astrometry of our catalog by exploiting new deep observations carried with the Wide-field InfraRed Camera (WIRCam) at the Canada–France–Hawaii telescope (CFHT) in the Ks-band, covering a $24' \times 24'$ area centered on the J1030 QSO (PI: M. Mignoli). From the radio and Ks-band catalogs cross-identification we found a matching rate of ${\sim}98$\%. The analysis of the separation between the radio and Ks-band position revealed a systematic offset in right ascension ($\Delta \mathrm{RA_{Ks-radio}} = -0.199$ arcsec), while no significant offset is found in declination ($\Delta \mathrm{DEC_{Ks-radio}} = -0.0005$  arcsec). 
The CFHT Ks-band observations are referred to the extremely precise \textit{Gaia} astrometry system \citep{gaia_collab}, which is in line with the VLBI reference frame. We therefore expected no significant offset between our and the Ks band  astrometry.  We verified that the observed RA offset was introduced during self-calibration, despite a clear overall improvement of the radio image (i.e, reduced overall noise level, reduced phase errors around the bright sources, better recovery of source low surface brightness emission). We hence decided to shift the image and catalog astrometry by the observed RA offset a posteriori. After this correction we measure $\Delta \mathrm{RA_{Ks-radio}} = -0.0042$ arcsec, which is negligible.

\subsection{Source sizes}
\label{sec:deconv}
\begin{figure}[t]
        \centering
\resizebox{\hsize}{!}
{\includegraphics[]{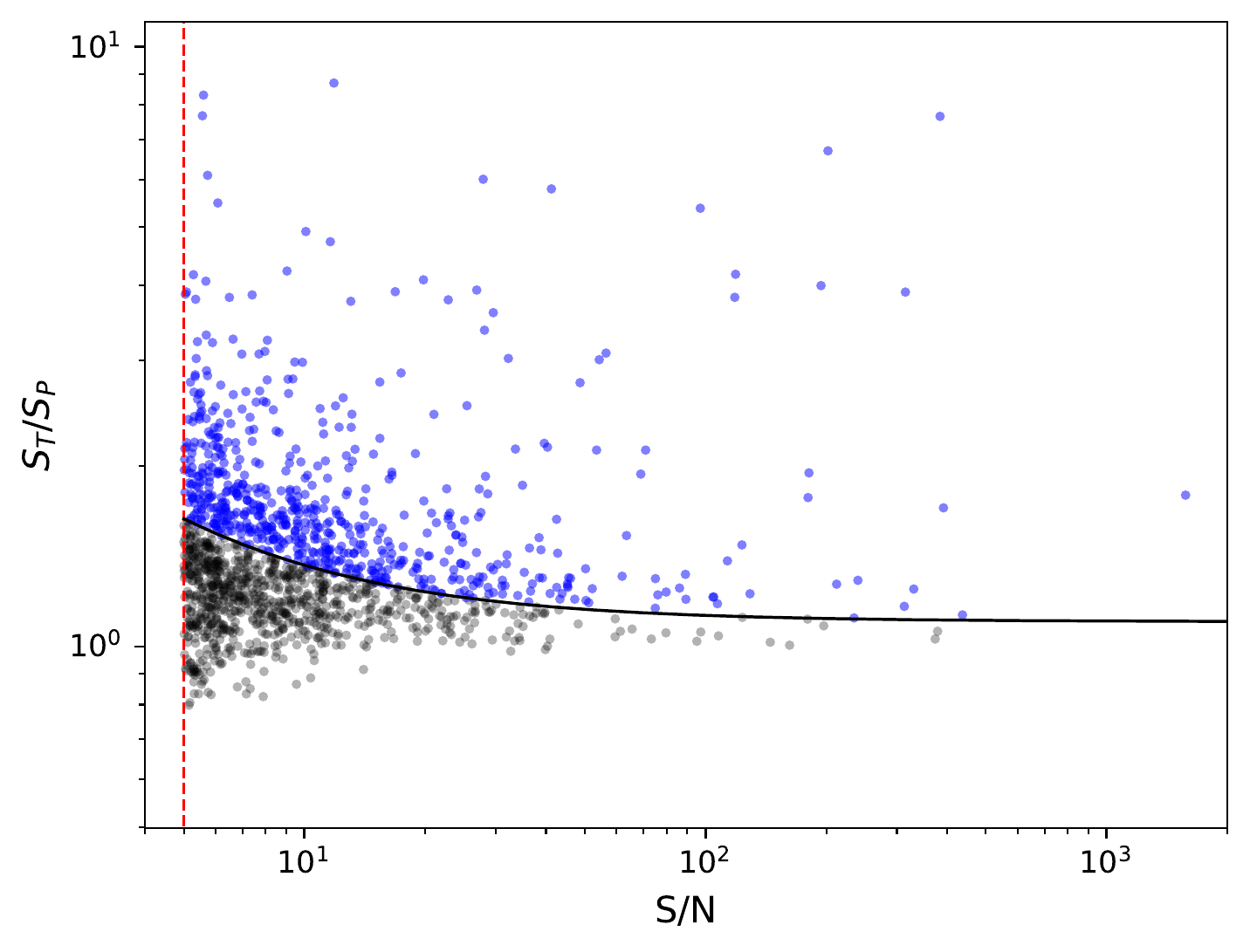}}

        \caption{Ratio between the total flux ($S_T$) and peak flux ($S_P$) for the sources of our catalog, as a function of S/N. The red vertical dashed line marks the detection threshold (S/N = 5). The black solid line (see Eq.~\ref{eq:res_func_num}) divides the resolved sources (above the line, blue points) from the unresolved sources (below the line, black points).}
        
        \label{fig:resolution}
\end{figure}

The ratio between the total flux density ($S_T$) and the peak flux density ($S_P$) of a radio source is inherently related to its extension, and in case of 2D-Gaussian fitting,  can be described by the following equation \citep{prandoni_2000}:
\begin{equation}\label{eq:res}
\frac{S_T}{S_P} = \frac{\theta_{\mathrm{min}}\theta_{\mathrm{maj}}}{ b_{\mathrm{min}} b_{\mathrm{maj}}}
\end{equation}
where $\theta_{\mathrm{min}}$ and $\theta_{\mathrm{maj}}$ are the minor and major axis of the source, respectively, and  $b_{\mathrm{min}}$ and $b_{\mathrm{maj}}$ are the minor and major axis of the synthesized beam, respectively. In Fig. \ref{fig:resolution} we report $S_T/S_P$ of our catalog as a function of S/N. We note that at low S/N there are many sources that, due to noise scattering, have $S_T/S_P$ $<$ 1. On the other hand, at high S/N random errors become negligible and systematic errors show up. In particular we see that  the ratio $S_T/S_P$ tends to ${\sim}1.1$, as expected, given the estimated 10\% radial smearing effect discussed in Sec. \ref{sec:smearing}. 

We can use the $S_T/S_P$  distribution as a function of S/N to discriminate between resolved and unresolved sources, by taking into proper account both random and systematic flux measurement errors. The minimum ratio $S_T/S_P$ that reliably separates resolved from unresolved sources can be expressed as a function of S/N by the following relation \cite[e.g.,][]{prandoni_2000,retana-monetenegro_2018,mandal_2021}:
\begin{equation}\label{eq:res_func}
\frac{S_T}{S_P} = \epsilon~\left[1+N\times \sigma\left(\frac{S_T}{S_P}\right)\right]
\end{equation}
where $\epsilon$ accounts for systematic errors and is set to 1.1, as found for high S/N sources. The factor $N \times \sigma(S_T/S_P)$ accounts for random errors and is expressed in number ($N$) of standard deviations $\sigma(S_T/S_P)$. At 90\% significance level N$\approx$1.64. In the case of 2D-Gaussian fits and Gaussian errors, \cite{condon_1997} derived general expressions for the relative errors of the fit parameters ($S_T$, $\theta_{\mathrm{maj}}$, $\theta_{\mathrm{min}}$) as a function of S/N, for unresolved sources.  In particular:
\begin{equation}\label{eq:axis_err_maj}
\frac{\sigma(\theta_{\mathrm{maj}})}{\theta_{\mathrm{maj}}} = k_1 ~\mathrm{S/N}^{-1}
\end{equation}
\begin{equation}\label{eq:axis_err_min}
\frac{\sigma(\theta_{\mathrm{min}})}{\theta_{\mathrm{min}}} = k_2 ~\mathrm{S/N}^{-1}
\end{equation}
where $k_1$ and $k_2$ are proportional factors. By fitting $\sigma(\theta_{\mathrm{maj}})/\theta_{\mathrm{maj}}$ and $\sigma(\theta_{\mathrm{min}})/\theta_{\mathrm{min}}$ as a function of S/N for our catalog, we found that $k_1$ and $k_2$ are equal to 1.16 and 0.91, respectively. By applying error propagation to  Eq. \ref{eq:res}, we can derive the relative error expression for the ratio $S_T/S_P$:
\begin{equation}\label{eq:ratio_err}
\sigma \left(\frac{S_T}{S_P}\right) = [k_1^2 + k_2^2]^{1/2} ~\mathrm{S/N}^{-1}\approx 1.47 ~\mathrm{S/N}^{-1} \, ,
\end{equation}
and, by substituting Eq. \ref{eq:ratio_err} into Eq. \ref{eq:res_func}, we find:
\begin{equation}\label{eq:res_func_num}
\frac{S_T}{S_P} = 1.1 \times \left[1+\left(\frac{2.42}{S/N}\right)\right] \, .
\end{equation}
This is the curve that subtends the unresolved source distribution at 90\% significance level (black solid line in Fig. \ref{fig:resolution}). Based on this analysis we can conclude that only 43\% of the catalogd sources can be reliably considered resolved. 
The resolved sources in our sample can be exploited to analyse the intrinsic source size distribution of $\mu$Jy radio sources (see Sec.~\ref{sec:res_bias}).

\subsection{Resolution bias}
\label{sec:res_bias}

\begin{figure}[h!]
        \centering
\resizebox{\hsize}{!}
{\includegraphics[]{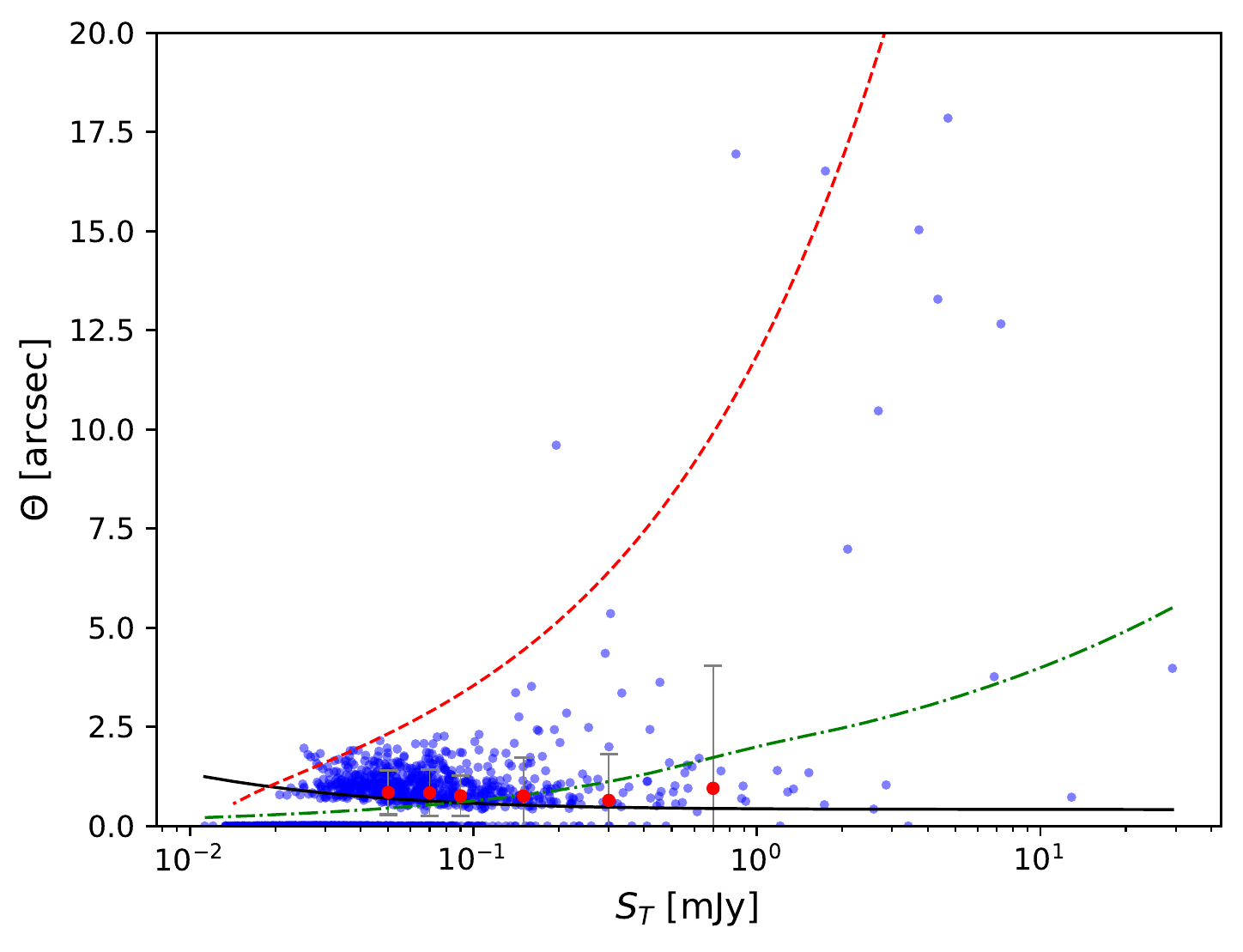}}

        \caption{Deconvolved source sizes as a function of total flux density. Unresolved source (as defined by Eq. \ref{eq:res_func_num}) have deconvolved sizes set to 0. The red circles indicate the median deconvolved sizes in bins of $S_T$ with 1$\sigma$ error indicated by the gray bars. The red dashed line is the maximum deconvolved size ($\Theta_{\mathrm{max}}$, Eq. \ref{eq:size_max}), the green dash-dotted line is the median size relation ($\Theta_{\mathrm{med}}$) derived from Eq. \ref{eq:size_med} and Eq. \ref{eq:mvar}, and the solid black line is the minimum deconvolved size ($\Theta_{\mathrm{min}}$, Eq. \ref{eq:size_min}).}
        
        \label{fig:median_size}
\end{figure}

The completeness of the catalog does not only depends on the S/N of the sources, but also on their angular size. Given a total flux density, the larger the source the wider the area over which the flux is distributed and the lower its peak flux density. Thus, a larger source will drop below the detection threshold more easily than a smaller source with the same total flux density. This is called resolution bias. In case of Gaussian errors, the resolution bias can be analytically treated following the procedure of \citeauthor{prandoni_2001} (\citeyear{prandoni_2001}; see also \citeauthor{mandal_2021} \citeyear{mandal_2021}). The maximum deconvolved size that a source can have without falling below the detection threshold of S/N = 5, as a function of the total flux density $S_T$, can be written as:
\begin{equation}\label{eq:size_max}
\Theta_{\mathrm{max}} = \Theta_{\mathrm{N}} \sqrt{S_T/(5\sigma) -1}
\end{equation}
where $\Theta_{\mathrm{N}}$ is the geometric mean of the synthesized beam axes and $\sigma$ is the image noise, that we set here equal to the median value of 3.4 $\mu$Jy (see Fig.~\ref{fig:visarea}). Exploiting Eq. \ref{eq:res} and Eq. \ref{eq:res_func_num} we can also derive an approximate function for the minimum deconvolved size (i.e., minimum size that can be resolved):
\begin{equation}\label{eq:size_min}
\Theta_{\mathrm{min}} = \Theta_{\mathrm{N}} \sqrt{1.1\times \left(1+\frac{2.42}{S_T/(5\sigma)}\right) -1} .
\end{equation}
In Fig. \ref{fig:median_size} we report the deconvolved sizes of our sources as a function of their total flux, rescaled to 1.4 GHz assuming a radio spectral index $\alpha=-0.7$. The deconvolved size is assumed equal to the geometric mean $\sqrt{\theta_{\mathrm{maj}} \theta_{\mathrm{min}}}$ if $\theta_{\mathrm{maj}}/\theta_{\mathrm{min}} <3$ and equal to $\theta_{\mathrm{maj}}$ otherwise. For sources which result unresolved based on Eq. \ref{eq:res_func_num}, we assume deconvolved sizes equal to 0. The red circles indicate the median deconvolved size of the sources in bins of $S_T$ from 0.04 mJy to 1 mJy. The bin size varies from 0.02 mJy to 0.6 mJy, in order to have sufficient statistics in each bin. We also plot the maximum ($\Theta_{\mathrm{max}}$) and minimum ($\Theta_{\mathrm{min}}$) deconvolved size, according to Eq. \ref{eq:size_max} and Eq. \ref{eq:size_min} (red dashed and solid black lines respectively). We note that the majority of the sources lie below the $\Theta_{\mathrm{max}}$ relation, as expected.

\begin{figure*}[h!]
        \centering
\resizebox{\hsize}{!}
{\includegraphics[]{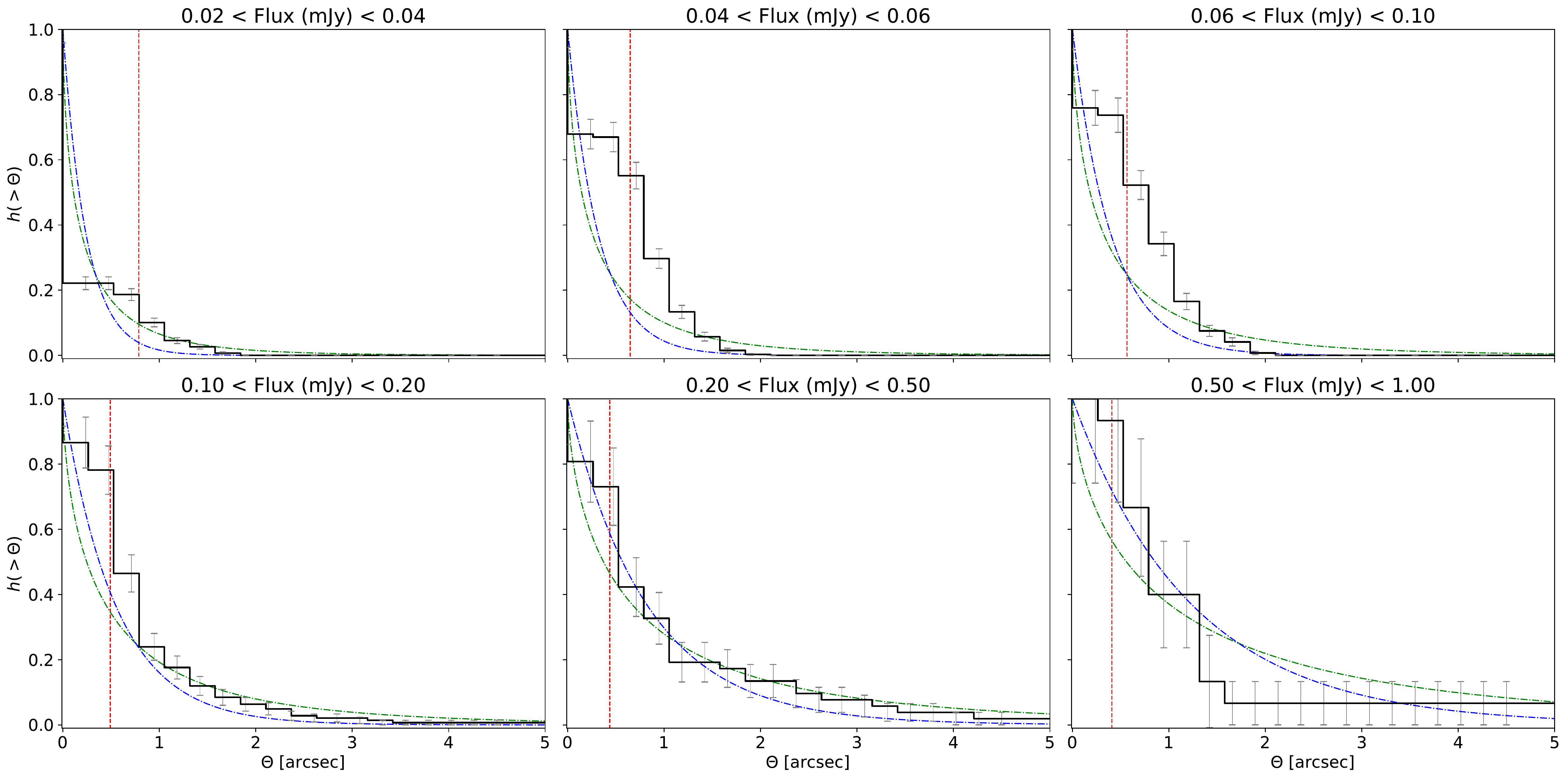}}

        \caption{Cumulative size distribution of our sources (black solid line) in six flux density bins, with 1$\sigma$ error indicated by the gray bars. The red dashed line is the minimum deconvolved size (Eq.~\ref{eq:size_min}), while the green and blue dash-dotted lines are the cumulative distribution functions described by Eq. \ref{eq:size_intdistr}, obtained by assuming $q=0.62$ and $q=1$ respectively. All lines are calculated at the geometric mean of each flux bin.}
        
\label{fig:size_int}
\end{figure*}

To correct for the resolution bias we need to infer the intrinsic size distribution of the sources; this will provide us with the number of sources, with a given flux density, that have a deconvolved size larger than $\Theta_{\mathrm{max}}$ and thus are missed in our survey. 
\cite{windhorst_1990} proposed the following (empirical) integral distribution of source sizes at 1.4 GHz, that has been subsequently used in several works to correct for resolution bias \cite[e.g.,][]{prandoni_2001, huynh_2005,hales_2014b, mandal_2021}:
\begin{equation}\label{eq:size_intdistr}
h(>\Theta)= \exp[-\ln 2 (\Theta/\Theta_\mathrm{med})^q]
\end{equation}
where $q=0.62$ and $\Theta_\mathrm{med}=\Theta_\mathrm{med}(S)$ is a function that describes how the source median size varies  as a function of flux density:
\begin{equation}\label{eq:size_med}
\Theta_\mathrm{med} = k \times (S_\mathrm{1.4~GHz})^m
\end{equation}
with $k=2$ arcsec, $m=3$, and $S_\mathrm{1.4~GHz}$ is in units of mJy. These parameters are best suited to describe source sizes down to 1.4 GHz flux densities of a few mJy. More recently, a steeper exponent $m$ was proposed for sub-mJy sources (\citealt{richards_2000,bondi_2008,smolcic_2017}). Following \cite{mandal_2021}, we assume a flux-dependent $m = m(S)$ exponent, to account for a smooth transition between the sub-mJy regime dominated by SFGs or host-confined AGN ($m = 0.5$) and the mJy regime dominated by extended radio galaxies ($m = 0.3$):
\begin{equation}\label{eq:mvar}
m(S_\mathrm{1.4~GHz}) = 0.3 + 0.2 \exp (-S_\mathrm{1.4~GHz}^2) .
\end{equation}
We find that the median sizes of our sample (red circles in Fig. \ref{fig:median_size}) are well reproduced by a $\Theta_\mathrm{med}$ relation with variable $m(S_\mathrm{1.4~GHz})$, as described by Eq.~\ref{eq:mvar} (see green dash-dotted line in Fig.~\ref{fig:median_size}). 

In Fig. \ref{fig:size_int} we show the integrated distribution size in bins of flux densities for our catalog. The red vertical lines mark the threshold below which the sources are considered unresolved in each flux bin (based on Eq. \ref{eq:size_min}) and the observed size distributions are not meaningful. So in the following we limit our analysis to the size bins above the aforementioned thresholds. \cite{mandal_2021} explored several values for the slope parameter ($q$) in Eq. \ref{eq:size_intdistr}. In particular, they found that the integral size distribution of their  low-frequency (150 MHz) sample is well described by a steeper function with $q=0.8$. We followed a similar approach, and  explored the slope range $q=0.62-1$, finding that the integral size distributions of our sources are fairly described by slopes within this range, as shown by the green and blue dash-dotted line in Fig. \ref{fig:size_int}. The only significant exceptions (deviation $\geq 3 \sigma$) consist of a couple of size bins in the flux intervals 0.04 $< S_T <$ 0.06 and  0.06 $< S_T <$ 0.1. Such deviations may indicate that different functional forms
should be explored, but they do not provide us with enough 
observational constraints to do so. We therefore decided to limit our analysis to the exponential functional form for consistent comparison with the existing literature. In addition, to account for resolution bias in the derivation of the source counts (Sec. \ref{sec:num_counts}), we decided to assume Eq. \ref{eq:size_intdistr} with $q=1$, and factored the uncertainties in the size distribution slope (down to $q=0.62$) into systematic error terms. As shown by \cite{prandoni_2001} the weight to be applied to source counts to correct for resolution bias incompleteness can be written as:
\begin{equation}\label{eq:cres}
C_\mathrm{res} = 1 / [1-h(>\Theta_\mathrm{lim})]
\end{equation}
where $\Theta_\mathrm{lim}$ is defined as $\Theta_\mathrm{lim}=\max[\Theta_\mathrm{min},\Theta_\mathrm{max}]$.

\subsection{Eddington bias}
\label{sec:eddbias}
In the presence of a nonuniform flux density distribution, the source flux densities do not follow a pure Gaussian noise distribution at the detection threshold. In particular, if the true source distribution decreases with increasing flux density, the flux densities tend to be boosted and the probability to detect a source below the detection threshold is higher than the probability to miss a source above the threshold, artificially boosting the detection fraction. This is the so-called Eddington bias \citep{eddington_1913}. 
One approach to correct for the Eddington bias consists in rescaling the measured fluxes to their intrinsic values, before deriving the number counts. A maximum likelihood solution for the true source flux density can be defined as follows (see \citealt{hales_2014a} and references therein): 
\begin{equation}\label{eq:edd_bias_intr}
S_\mathrm{true} = \frac{S_\mathrm{meas}}{2} \left(1+ \sqrt{1-\frac{4\gamma}{\mathrm{S/N}^2}}\right)
\end{equation}
where $\gamma=\gamma(S)$ is the exponent of the intrinsic number count distribution at a given flux density, modeled by a power-law ($d$N/$d$S $\sim$ $S^{-\gamma}$). The slope of the counts can be modeled from empirical polynomial fits of the observed counts. There are several fits available in the literature. Following \cite{mandal_2021}, in this work we used as a fiducial model the sixth-order polynomial fit by \citeauthor{bondi_2008} (\citeyear{bondi_2008}; that is $\gamma=2.5$, consistent with an Euclidean model), and we factored the uncertainties associated to other fits into a systematic error term, when correcting the source counts for the Eddington bias (see \citealt{mandal_2021} for more details).

\section{Source counts}
\label{sec:num_counts}
\begin{figure*}[h!]
        \centering
{\includegraphics[scale=0.6]{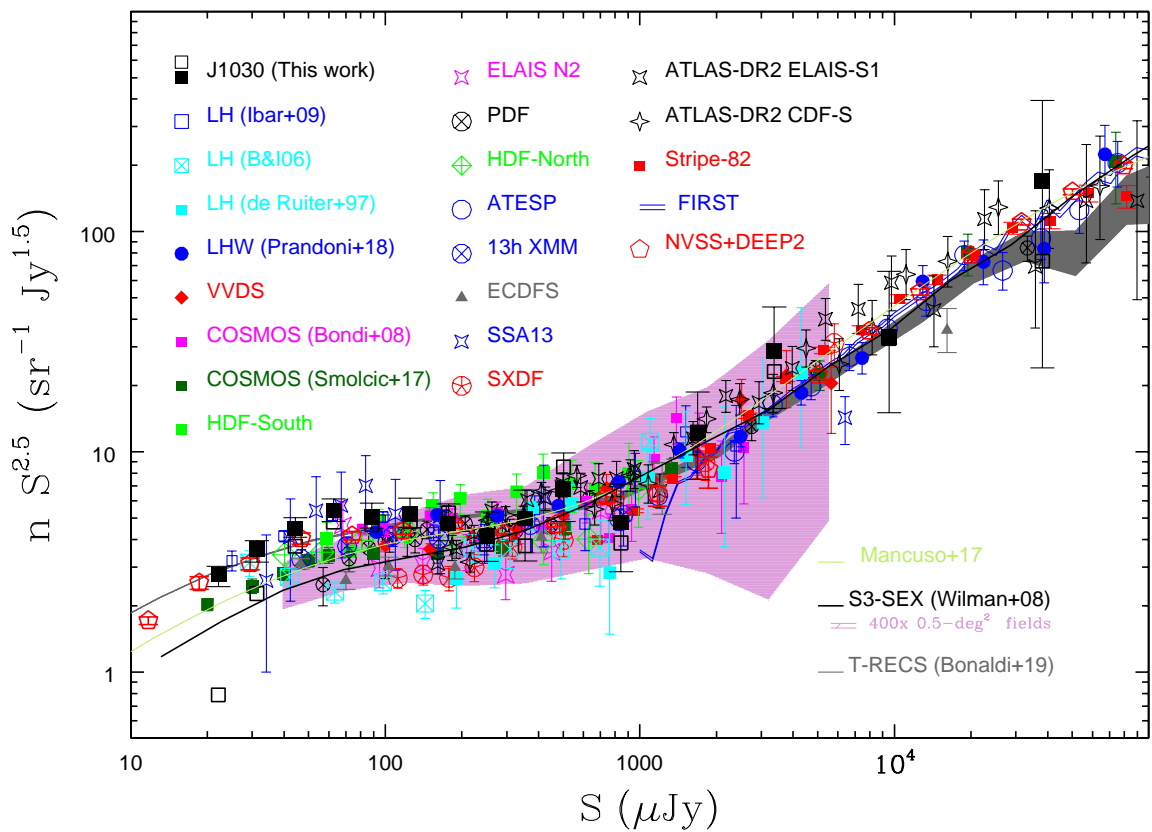}}
\vspace{-0.8cm}
        \caption{Normalized 1.4 GHz differential source counts for different samples (as indicated in the figure and in the text).
        The counts derived from the catalog discussed in this work are represented as filled black squares, and have been rescaled from 1.34 GHz to 1.4 GHz by assuming a spectral slope $\alpha=-0.7$. Vertical bars represent the squared sum of Poissonian and systematic errors on the normalized counts (see text for more details). The empty black squares show the same counts, before applying the corrections for the visibility function and for the systematic effects discussed in Sec. \ref{sec:cat}. We also show the source counts derived from the \citet[][S3-SEX]{wilman_2008}, \citet[T-RECS]{bonaldi_2019} and \citet{mancuso_2017} 1.4 GHz simulations, which represent the summed contribution of the modeling of various source populations (RL and RQ AGN; SFGs).  The predicted spread due to cosmic variance for 0.5 deg$^2$ fields is represented by the pink shaded area. It has been obtained by splitting the S3-SEX simulation in 400 0.5-deg$^2$ fields.}
        
        \label{fig-diffcounts}
\end{figure*}

The differential source counts normalized to a nonevolving Euclidean model ($S^{2.5}$) are  listed in Table~\ref{tab-counts} and shown in Figure~\ref{fig-diffcounts} (filled black squares).  
Each source has been weighted by the reciprocal of its visibility area. The counts were corrected for the systematic errors as described in Sect \ref{sec:cat}. 
The uncertainties associated with such corrections are factored into systematic error terms (see Sys$^-$ and Sys$^+$ columns in Table \ref{tab-counts}).

\begin{table*}
\caption{\label{tab-counts} 1.34 GHz source counts as derived from our survey.
}
\centering
\begin{tabular}{ccccccccc}
\hline \hline
$\Delta S$ [mJy] & $\langle S\rangle$ [mJy] & $N_{S}$ & $d$N/$d$S $\mathrm{S^{2.5}}$ [$\mathrm{sr^{-1} Jy^{1.5}}$] & $\sigma^-$ & $\sigma^+$ & $\rm{Sys}^-$ & $\mathrm{Sys}^+$ & $d\mathrm{N^{uncorr}}$/$d$S $\mathrm{S^{2.5}}$ [$\mathrm{sr^{-1} Jy^{1.5}}$] \\


(1) & (2) & (3) & (4) & (5) & (6) & (7) & (8) & (9) \\
\hline
      $0.019   -   0.027$   &    0.023   &  207   &  2.89   &  0.20   &  0.21   &  0.27   &  0.16  & 0.82  \\	
      $0.027   -   0.038$   &    0.032   &  297   &  3.79   &  0.22   &  0.23   &  0.43   &  0.26  &   2.38   \\	
      $0.038   -   0.054$   &    0.045   &  279   &  4.69   &  0.28   &  0.30   &  0.28  &  0.48   &  3.9  \\	
      $0.054   -   0.077$   &    0.065   &  209   &  5.64   &  0.39   &  0.42   &  0.44   &  0.63  &  5.03  \\   
      $0.077   -   0.109$   &    0.091   &  121   &  5.32   &  0.48   &  0.53   &  0.37   &  0.61  &  4.89  \\   
      $0.109   -   0.154$   &    0.129   &   73   &  5.48   &  0.64   &  0.72   &  0.37   &  0.66  &  5.05 \\	
      $0.154   -   0.217$   &    0.183   &   40   &  4.94   &  0.78   &  0.91   &  0.19   &  0.67  &  5.53 \\	
      $0.217   -   0.307$   &    0.258   &   21   &  4.34   &  0.95   &  1.15   &  0.34   &  0.50  &   3.99  \\	
      $0.307   -   0.435$   &    0.365   &   15   &  5.21   &  1.34   &  1.69  &  1.29   &  0.63   &  3.91 \\	
      $0.435   -   0.615$   &    0.517   &   12   &   7.14  &  2.06   &  2.66   &  0.36   &  1.80  &  8.93  \\	
      $0.615   -   1.229$   &    0.869   &    8   &   5.02  &  1.78   &  2.40    &  0.98   &  0.58 &  4.04  \\   
      $1.229    -  2.459$   &    1.739   &    7   &  12.68  &  4.79   &  6.60   &  0.67   &  1.96  &  12.9  \\	
      $2.459   -   4.918$   &    3.477   &    6   &  30.03  &  12.26   &  17.27  &  5.78   &  3.66 &  24.25  \\   
      $4.918   -   19.670$  &    9.835   &    3   &  34.82  &  18.92  &  33.89  &  2.17  &  4.87   &  34.29  \\   
      $19.670   -  78.680$  &    39.340  &    2   &  177.1  &  114.2 &  233.8 &  100.2  &  21.77   &  76.89   \\	
\hline    	  
\end{tabular}
\tablefoot{ (1) Flux density interval. (2) Geometric mean of the flux density. (3) Number of sources detected. (4) Differential counts normalized to a non evolving Euclidean model. (5) and (6) Poissonian errors on the normalized counts. (7) and (8) Systematic errors, accounting for different modeling of resolution and Eddington bias corrections (see Sects. \ref{sec:res_bias} and \ref{sec:eddbias} for more details). (9) Normalized differential counts before applying the corrections for the visibility function and for the systematic effects discussed in Sec. \ref{sec:cat}.}

\end{table*}

Our source counts are compared with 1.4 GHz counts available from literature. These include all known deep fields, from subdegree fields such as13 hours XMM--\textit{Newton}/ROSAT \citep{seymour_2004}, Lockman Hole \citep{deruiter_1997,biggs_2006,ibar_2009}, Hubble deep field north and south \citep{huynh_2005,biggs_2006}, European large area ISO survey N2 \citep{biggs_2006}, Subaru/XMM--\textit{Newton} deep field \citep{simpson_2006}, small selected area 13 \citep{fomalont_2006} and extended \textit{Chandra} deep field south \citep{padovani_2015}, to wider-area ($> 1~\mathrm{deg^2}$) surveys, such asPhoenix deep survey \citep{hopkins_2003}, VLA-VIRMOS VLT \citep{bondi_2003}, VLA cosmic evolution survey\cite[COSMOS][]{bondi_2008,smolcic_2017b}, Australia telescope large area survey \cite[ATLAS][]{hales_2014a}, the Westerbork Lockman hole mosaic \citep{prandoni_2018}, to shallower large ($\gg 10~\mathrm{deg^2}$) surveys such as the Australian telescope ESO slice project \cite[ATESP][]{prandoni_2001}, SDSS Stripe 82 \citep{heywood_2016} and FIRST \citep{white_1997}. We also show the deepest and most robust determination to date: the one derived by \citet{matthews_2021}, by combining the MeerKAT Deep2 field and the NVSS survey (red pentagons). Finally we show simulated source counts derived by combining evolutionary models of either classical RL AGN or faint radio source populations dominating the sub-mJy radio sky, namely SFGs and RQ AGN. In particular we show the 1.4 GHz counts derived from the modeling of \cite{mancuso_2017} (light green solid), the 25 $\mathrm{deg^2}$ tiered radio extragalactic continuum simulation \cite[T-RECS, dark gray shaded area,][]{bonaldi_2019}, as well as the source counts derived from the semi-empirical sky simulation developed in the framework of the SKA simulated skies project \cite[S3-SEX, black solid line,][]{wilman_2008,wilman_2010}.

Figure~\ref{fig-diffcounts} illustrates very well the long-standing issue of the large scatter (exceeding Poisson fluctuations) which is observed at flux densities $\la 1$ mJy. The main cause for this considerable scatter is the combination of cosmic variance \cite[][]{heywood_2013} and survey systematics introduced by calibration, deconvolution, source extraction algorithms, and corrections such as those discussed in Sect.~\ref{sec:cat} \cite[see][]{condon_2012}. These issues have been extensively discussed in the recent literature \citep[see, e.g.,][]{heywood_2016,prandoni_2018}. The role of cosmic variance is clearly illustrated by the pink shaded area shown in Fig. \ref{fig-diffcounts}, showing the predicted source counts' spread for small-area ($\ll$1 deg$^2$) deep radio fields. This has been obtained by splitting the S3-SEX simulation in 400 0.5-deg$^2$ fields.  
Fig. \ref{fig-diffcounts} clearly shows that our source counts are among the deepest available so far, and are overall consistent with both \citet{matthews_2021} source counts and the recent simulations by \citet{bonaldi_2019}. 
We note, however, that our survey possibly shows an excess of faint sources at ${\sim}$50 $\mu$Jy. The observed excess is fully consistent with sample variance, which at this flux density can be quantified in ${\sim}$10\% for a survey covering 0.2 deg$^2$, such as the present one \citep[see][]{heywood_2013}. Indeed, as discussed in Sec.~\ref{sec:intro}, J1030 is known to be a biased field, that possibly contains several overdensities \citep{marchesi_2021}. To date, the two confirmed ones are the one assembling around the $z=6.3$ QSO \citep{mignoli_2020}, and the protocluster at $z=1.7$  \citep{gilli_2019,damato_2020a}. We note that these structures are likely more extended and contain more objects that those confirmed to date \citep{balmaverde_2017, peca_2021}. A multiband analysis of the radio catalog presented in this work is currently in progress and will be presented in a future paper. As part of this future paper the photometric and spectroscopic redshifts will be derived, allowing us to further explore the hypothesis of cosmic variance being responsible for the observed source counts' excess.

\section{Notable sources}
\label{sec:pec_sources}

\begin{figure}[h!]
        \centering
\resizebox{\hsize}{!}
{\includegraphics{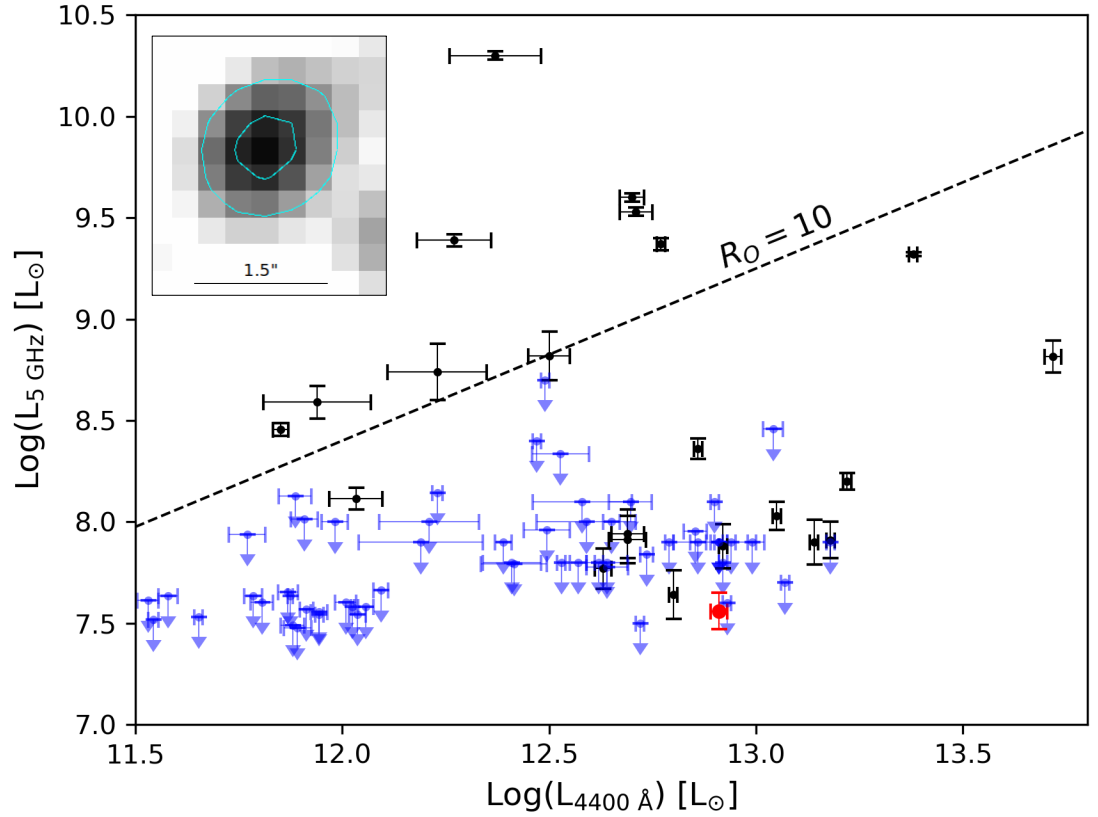}}

        \caption{Rest-frame optical luminosity at 4400 $\AA$ versus rest-frame radio luminosity at 5 GHz for known QSOs at $z{\gtrsim}6$ (\citeauthor{banados_2015} \citeyear{banados_2015} and references therein; \citeauthor{banados_2018} \citeyear{banados_2018}; \citeauthor{liu_2021} \citeyear{liu_2021}; \citeauthor{sbarrato_2021} \citeyear{sbarrato_2021}). Black points mark radio detected sources, while blue points indicate radio undetected ones, for which 3$\sigma$ upper limits are reported. The red point marks the position of the SDSS J1030+0524 QSO, based on  our JVLA  detection. The black dashed line indicates the radio loudness value $R_O = 10$, set as a threshold between RQ and RL AGN \citep{jiang_2007}. The inset shows a JVLA image cutout of the $z$ = 6.3 QSO, with cyan contours starting at 3$\sigma$ and increasing with a geometric progression of $\sqrt{3}$. The black bar indicates the angular scale of the image.
        }
        
        \label{fig:QSOz6}
\end{figure}

\begin{figure*}[h!]
        \centering
\resizebox{\hsize}{!}
{\includegraphics[]{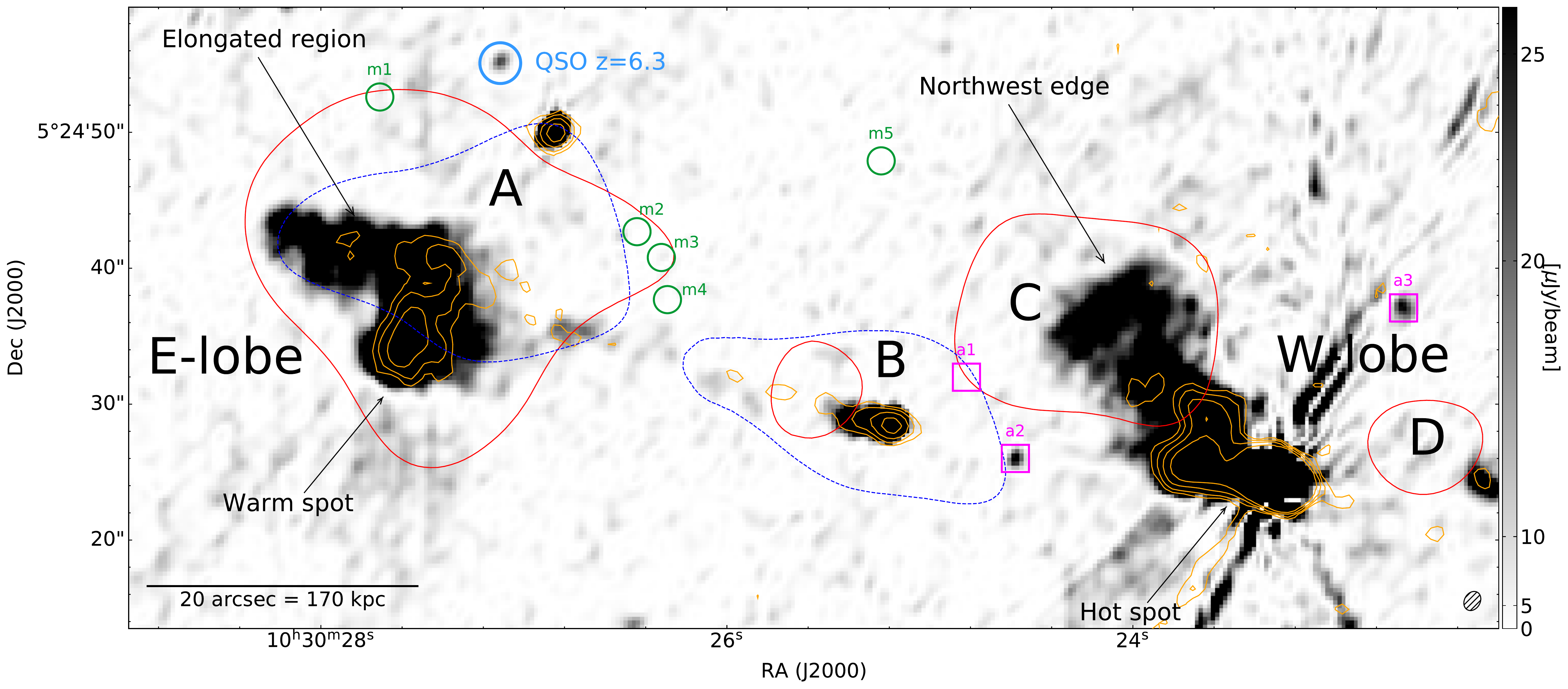}}

        \caption{Cut-out of the 1.34 GHz JVLA image centered on the HzRG at $z=1.7$. The VLA contours at 1.4 GHz \citep{petric_2003,nanni_2018} are reported in orange, starting from a 3$\sigma$ threshold and increasing with a $\sqrt{3}$ geometric progression. Blue and red contours are the ${\sim}$2$\sigma$ hard and soft X-ray diffuse emission, respectively. The four major components of such emission are labeled in black (A, B C and D). The positions of five of protocluster members discovered by MUSE ($m1-m5$) are indicated by the green circles (a sixth source is located outside the cut-out, ${\sim}$45 arcsec northeast from the FRII galaxy core). The positions of the protocluster members discovered by ALMA ($a1-a3$) are marked by the magenta boxes. The eastern and western lobes are labeled as E-lobe and W-lobe, respectively. The main morphological features of the lobes are also reported, indicated by the black arrows. Radial features originating from the western hot-spot are due to residual phase errors. The light blue circle indicates the $z=6.3$ QSO at the center of the J1030 field. The solid black line at the bottom-left corner indicates the angular and physical scale, while the dashed ellipse at the bottom-right corner is the restoring beam of the JVLA image.}
        
\label{fig:protoc}
\end{figure*}

\subsection{QSO SDSS J1030+0524 at $z=6.3$}
\label{subsec:QSOz6}

We detect, for the first time in the radio band, the $z=6.3$ SDSS J1030+0524 QSO at the field center (RID: 762, see inset in Fig. \ref{fig:QSOz6}). The measured flux density at the effective frequency is $S_{1.34~\mathrm{GHz,~obs}} = 26 \pm 5 ~\mu$Jy, corresponding to a 1.4 GHz rest-frame luminosity $L_{1.4~\mathrm{GHz,~rest}} = 6.2 \pm 1.2 \times 10^{24}$ W/Hz, assuming $\alpha=-0.7$. The previous 3$\sigma$ upper limit to the 1.4 GHz flux density, reported by \cite{petric_2003}, was 60 $\mu$Jy. On the basis of this constrain, \cite{banados_2015} calculated the 3$\sigma$ upper limit to the 5 GHz rest-frame integrated luminosity, that is $\log(L_{5~\mathrm{GHz,~rest}}/L_{\odot})~<~$  7.9. 
Furthermore, from the 3.6 $\mu$m flux density \cite[74 $\pm$ 3 $\mu$Jy,][]{leipski_2014}, \cite{banados_2015} derived the UV-rest-frame luminosity ($\log(L_{4400~\mathrm{\AA,~rest}}/L_{\odot})$ ${\sim}$ 12.91) assuming an optical spectral index of $-0.5$. On the basis of the UV-rest-frame flux density $f_{4400~\AA}$ and the upper limit to the 5 GHz rest-frame flux density $f_{5~\mathrm{GHz}}$, \cite{banados_2015} derived an upper limit for the optical radio loudness $R_O = f_{5~\mathrm{GHz}}/f_{4400~\AA} < 1.5$.
This upper limit is well below the $R_O$ = 10 threshold, that marks the division between RQ and RL AGN, following the definition by \cite{jiang_2007}. Thanks to our detection we can now securely estimate the optical radio loudness of this object: from $S_{1.34~\mathrm{GHz,~obs}}$ we derived a rest-frame $\log(L_{5~\mathrm{GHz,~rest}}/L_{\odot})$ = 7.51 $\pm$ 0.09 and $R_O =$ 0.62 $\pm$ 0.12. 
Thus, our measurement reveals that SDSS J1030+0524 is the faintest radio detected QSO at $z\gtrsim6$ discovered so far and the most RQ (see Fig. \ref{fig:QSOz6}). However, we note that J1030+0524 may not be the most radio quiet AGN, but just the most radio quiet among those radio detected so far. Indeed there are undetected sources whose 3$\sigma$ radio upper limits point toward similar radio/optical ratios, or even lower. It is interesting to note that this RQ AGN is at the center of a giant assembling structure \citep{mignoli_2020}, the most distant known to date.
Several works have shown that high-$z$ ($z\gtrsim2$) protoclusters preferentially assemble around powerful RL AGN \cite[e.g.,][]{pentericci_1997,venemans_2007, miley_2008, overzier_2016,gilli_2019}, possibly up to $z\sim5.2-5.8$ \citep{overzier_2006,zheng_2006}. In addition, \cite{liu_2021} have shown that the radio loud fraction (RLF) of $z{\sim}6$ radio galaxies is consistent with the low-redshift fraction (${\sim}$10\%), suggesting no evolution of the RLF for optically selected QSOs at different redshifts. 
However the available statistics is sparse and the link between radio loudness and high-$z$ overdensities is still an open issue. Further wide-field deep observations around high-$z$ QSOs are needed to reach firmer conclusions. 

\subsection{FRII galaxy at $z=1.7$: a proto-BCG}

The FRII galaxy (RID: 753), around which the $z$ = 1.7 protocluster is assembling, is the source with the largest angular size in the field, and the second brightest in our sample. Its measured flux density is $S_{1.34~\mathrm{GHz,~obs}} = 25 \pm 4$ mJy. By assuming $\alpha = -0.7$, this measurement corresponds to $S_{1.4~\mathrm{GHz,~obs}} = 24 \pm 5$ mJy, consistent with the NVSS value (${\sim}$30 mJy) and with that measured by \cite{nanni_2018} in the old VLA observations (${\sim}$27 mJy). Its 1.4 GHz luminosity is $(3.5 \pm 0.7) \times 10^{26}$ W Hz$^{-1}$, which is typical of FRII radio galaxies.

The FRII galaxy is shown in Fig. \ref{fig:protoc}. The new JVLA observations presented in this work reveal significant (${>}$3$\sigma$) more extended emission in both the lobes of the FRII at the center of the $z$ = 1.7 protocluster, with respect to the previous observations reported by \citeauthor{petric_2003} (\citeyear{petric_2003}; orange contours, see Sec. \ref{sec:intro}). The W-lobe of the radio galaxy features a clear hotspot, whose emission constitutes ${\sim}$75\% of the total source flux density. In the E-lobe a more complex geometry is unveiled, with the lack of a classical hotspot \cite[i.e., brightness contrast with the rest of the radio source $\geq4$, following the definition of][]{deruiter_1990}. The brightest region of the E-lobe is a ``warm spot'' located south of the lobe. The new JVLA observations reveal the presence of diffuse emission located northeast from the warm spot (at a distance of ${\sim}$5 arcsec), and elongated in the east-west direction. Substantial extended emission is also revealed by the JVLA northeast of the western hot-spot. 

As discussed in Sec.~\ref{sec:intro}, diffuse X-ray emission has been detected around the FRII radio galaxy. The several spots of diffuse X-ray emission (named A, B, C and D) are indicated in Fig.~\ref{fig:protoc} by red and blue contours, for soft and hard band emission respectively. The origin of this emission is  discussed in  \cite{gilli_2019}. The A spot, overlapping with the E-lobe of the radio galaxy, likely originates from a combination of both thermal and nonthermal (Inverse Compton scattering between the Cosmic Microwave Background and the FRII galaxy lobe electrons, IC-CMB) processes. The presence of nonthermal emission is supported by the apparent overlap between the hard X-ray emission and the new east-west elongated feature revealed by the JVLA. 
The B spot is mostly hard and coincides with the FRII galaxy nucleus and jet base; its emission is well described by nonthermal processes arising from the inner part of the jet. Instead, the soft band C component is well fitted by a thermal emission model. The C spot is particularly interesting as it seems to overlap with the W-lobe extended emission revealed by the JVLA, suggesting a possible interaction between the ICM and the FRII radio galaxy. This seems to be supported by an apparent enhancement of the radio emission at the northwest edge, possibly associated with turbulence or shocks. 

\cite{gilli_2019} provided and estimate of the FRII galaxy inclination angle ($70^\circ - 80^\circ$), based on the jet versus counter-jet base flux density ratio $R_\mathrm{jet}$. Thanks to the higher resolution and sensitivity of the new observations, the core and the eastern jet base are better resolved in the JVLA image. Thus, we can provide a more reliable estimate of the FRII galaxy inclination angle. As in \cite{gilli_2019}, we assume a jet velocity $\beta = 0.5$, a Doppler boost exponent $p=2$, and a jet base spectral index $\alpha_\mathrm{jet}=-0.5$, which is typical for local radio galaxies and one-sided FRII galaxies \citep{giovannini_2001,ruffa_2019}. For a given $R_\mathrm{jet}$
the inclination $\theta$ can be derived from the following equation \citep{giovannini_1998}:
\begin{equation}\label{eq:Rjet}
R_\mathrm{jet} = [(1+k)/(1-k)]^{p+\alpha_\mathrm{jet}}, ~\mathrm{where}~k=\beta\cos({\theta}).
\end{equation}
We measure $R_\mathrm{jet}{\sim}2$, from which we derive  $k{\sim}0.14$, corresponding to $\theta{\sim}80^\circ$, consistent with \cite{gilli_2019} findings. We hence confirm that the radio galaxy is almost lying in the plane of the sky. 
We also provide a second estimate of the inclination angle, based on the jet length ratio $L_\mathrm{jet}/L_\mathrm{cjet}$, again following \cite{giovannini_1998}:
\begin{equation}\label{eq:Ljet}
L_\mathrm{jet}/L_\mathrm{cjet} = (1+k)/(1-k).
\end{equation}
In this latter case we assume two scenarios:  the approaching jet ends at a) at the end of the elongated region revealed by the JVLA in the E-lobe or b) at the warm spot. 
In the former case we find $k{\sim}0.25$, in the latter we obtain $k{\sim}0.14$, nicely in agreement with the value obtained from $R_\mathrm{jet}$. This suggests that the warm spot may indeed correspond to the end of the approaching eastern jet, which would likely imply a bending of the jet. This interpretation is strengthened by the analysis of the polarized emission of the FRII radio galaxy presented by  \cite{damato_2021}: the magnetic field shows a wrapping around the peak emission of the total intensity in the warm spot, as observed in several cases around FRII galaxies' hot spots.

Finally, we provide a revised estimate of the source size, based on the full extent of the radio emission. The radio galaxy (deconvolved) major axis is measured to be 1.3 arcmin, which corresponds to a projection-corrected physical size of ${\sim}$700 kpc (to be compared with the value of ${\sim}$600 kpc reported by \citealt{gilli_2019}). This makes this source a giant radio galaxy.

\subsection{A possible second radio AGN in the protocluster}

The gas-rich protocluster members $a2$ and $a3$, discovered by \cite{damato_2020b}, are both detected in the JVLA image (RID=813 and RID=755). The sources are unresolved and their flux densities are $S_{1.34~\mathrm{GHz,~obs}} = 24 \pm 5$ $\mu$Jy  and $S_{1.34~\mathrm{GHz,~obs}} = 30 \pm 9$ $\mu$Jy, respectively. The latter source, however, is associated with a blend of two optical  galaxies \citep{damato_2020b}, and hence its measured parameters cannot be considered fully reliable. Focusing on $a2$, and 
assuming that its radio emission is fully ascribed to star-formation activity, we can derive a radio-based SFR. This is done by exploiting the relation presented by \cite{novak_2017}, and results in SFR${\sim}90-150~\mathrm{M_\odot}$/yr, depending on whether a Salpeter or a Chabrier initial mass function (IMF) is assumed \citep{salpeter_1955,chabrier_2003}. This SFR range is ${\sim}2-4$ $\times$ the SFR of the source measured through the Schmidt-Kennicutt (SK) law by \cite{damato_2020b}, that is ${\sim}40-60~\mathrm{M_\odot}$/yr. We derived another independent SFR estimate for $a2$, by exploiting the broad-band coverage of the J1030 field, and performing SED fitting. In Table \ref{tab:a2_SED} we report the photometry of the source in each available optical and IR band, the observing instrument and the photometry references. 
We used the \textit{Hyperz} code \citep{bolozonella_2000} to fit the SED of $a2$, deriving a stellar mass $\log(M_{\ast}/\mathrm{M_{\odot}}) = 10.75 \pm 0.25$ and an upper limit to the SFR (at the 90\% significance level) of 43 $\mathrm{M_{\odot}}$/yr. The derived specific SFR upper limit is sSFR${<}7\times 10^{-10}$ $\mathrm{yr^{-1}}$, which is comparable with the sSFR of normal star-forming galaxies at the same redshift \citep{schreiber_2015}. The SFR obtained from SED fitting is in good agreement with the lower end of the SFR range derived by \cite{damato_2020b} on the basis of the cold molecular gas emission, and is at least a factor 2 lower than the radio-based SFR. This seems to suggest that at least part of the observed radio emission is of AGN origin. If confirmed, $a2$ would be the second radio AGN detected in the protocluster, located closely to the assembling proto-BCG (projected distance ${\sim}80$ kpc).

\begin{table}
\caption{\label{tab:a2_SED} Optical/IR photometry for source $a2$.
}
\centering
\resizebox{\hsize}{!}{
\begin{tabular}{cccc}
\hline \hline
Band & $M_{AB}$ & Instrument & Reference \\
(1) & (2) & (3) & (4) \\
\hline
$r$ & $>$26 & LBT/LBC & \cite{morselli_2014}\\
$i$ & $>$25.5 & LBT/LBC &    ''\\
$z$ & $26.6 \pm 0.2$ & HST/ACS & \cite{stiavelli_2005}\\
$Y$ & $25.4 \pm 0.5$ & CFHT/WIRCam & \cite{balmaverde_2017}\\
$J$ & $23.7 \pm 0.2$ & CFHT/WIRCam & ''\\
$H$ & $23.5 \pm 0.2$ & HST/WFC3 & \cite{damato_2020b}\\
$K$ & $22.6 \pm 0.1$ & CTIO/ISPI & \cite{quadri_2007}\\
CH1  & $20.8 \pm 0.7 $ & Spitzer/IRAC   &  \cite{annunziatella_2018} \\
CH2  & $20.6 \pm 0.5$ & Spitzer/IRAC   & ''\\

\hline    	  
\end{tabular}
}
\tablefoot{(1) Observed band. (2) Measured $AB$ magnitude 
(careful analysis was required due to contamination from a nearby bright star). The lower limits are given at the 5$\sigma$ level. (3) Observing instrument. (4) photometry reference. 
}

\end{table}

\section{Radio/X-ray luminosity correlations}
\label{sec:X_rays}
As discussed in Sec.~\ref{sec:intro}, the J1030 field has been the target of deep (${\sim}$500 ks) \textit{Chandra} observations. The region covered by \textit{Chandra} is fully included in the radio field (see cyan solid line in Fig. \ref{fig:field}). \cite{nanni_2020} extracted a catalog of 256 X-ray sources from these observations. By exploiting the multiwavelength coverage of the field, \cite{marchesi_2021} derived photometric redshifts for 243 of the 256 X-ray sources, through SED fitting. For 123 of them, they also obtained spectroscopic redshifts, mostly through observations with the LBT. The optical spectra also allowed them to derive a spectral classification for these sources, as follows: 43 sources are classified as broad-line AGN (BL-AGN), 20 are narrow-line AGN (NL-AGN), 28 are emission line galaxies (ELG) and 32 are early-type galaxies (ETG). We notice that the ELG and ETG classes may be contaminated by AGN, if AGN spectral features fall outside the spectral range sampled by the observations or are not visible due to obscuration  \citep{marchesi_2021}.

\begin{figure}[t]
        \centering
\resizebox{\hsize}{!}
{\includegraphics[]{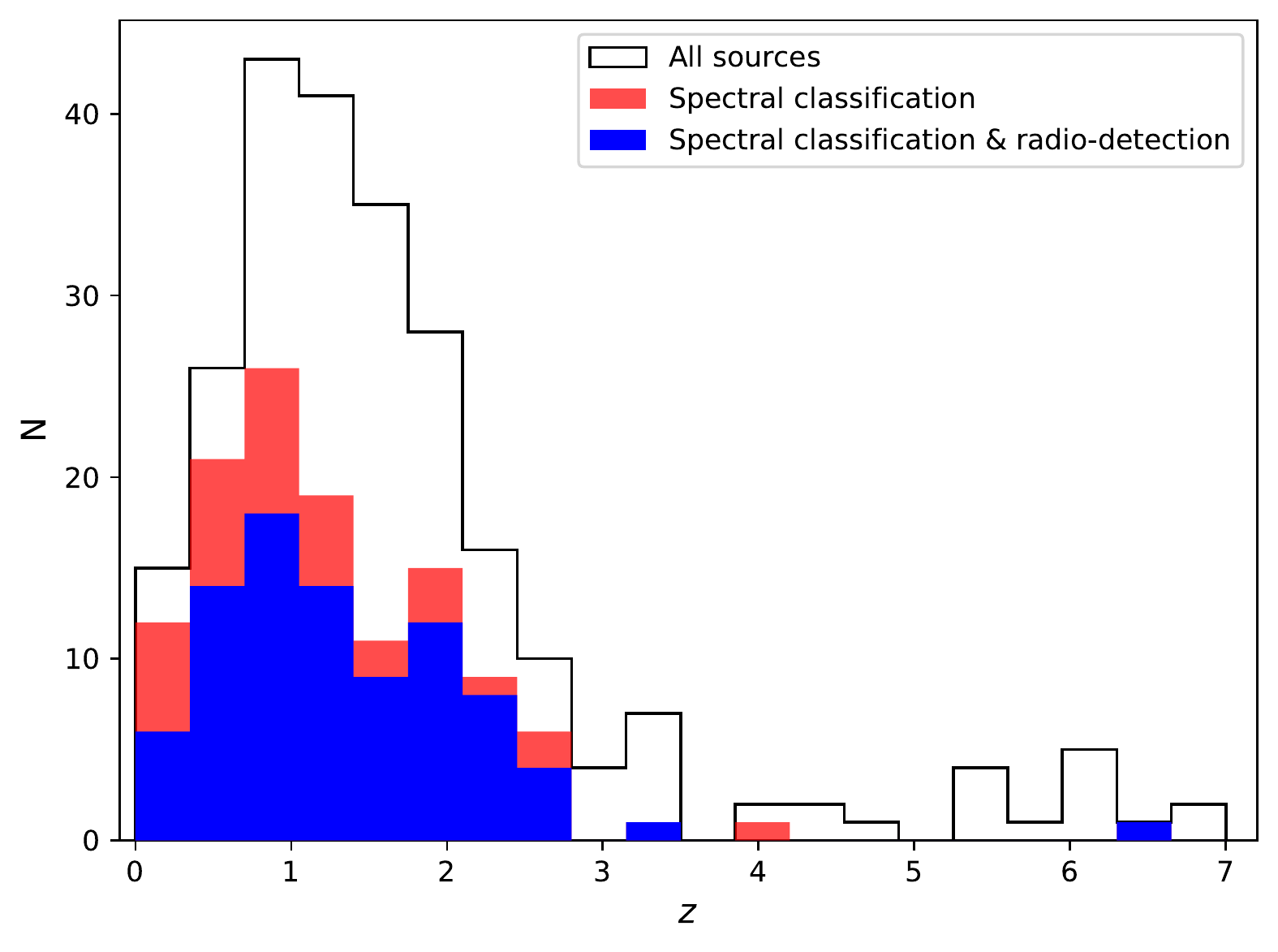}}

        \caption{Redshift distribution of the 243 X-ray sources (black solid line) analyzed by \cite{marchesi_2021}, in redshift bins of 0.35. The distribution of the 123 sources with spectroscopic redshift is indicated by the filled red histogram. The filled blue histogram highlights the distribution of those, among the latter, with a radio counterpart.}
        
        \label{fig:red_distr}
\end{figure}

The 243 X-ray sources with photometric and/or spectroscopic redshifts can be exploited to investigate AGN X-ray/radio luminosity correlations and their dependence on source type and redshift. To this end, we matched  the 243 X-ray sources analyzed by \cite{marchesi_2021} with our radio catalog. To do so, we exploited the positions for the optical/NIR counterparts of the X-ray sources (provided by \citealt{nanni_2020}), as they are more accurate than X-ray positions. We used a matching radius of 1.5 arcsec, which appears appropriate from the analysis of the optical/radio position separation distribution. We note that the positional error on both the optical and radio position is ${>}$10${\times}$ lower than the chosen matching radius, and that we visually inspected all matches to identify possible spurious associations. As a result, we found 104 reliable matches. For the 139 X-ray sources with no radio counterpart in the catalog, we searched for additional radio detections by performing a parabolic fitting of the radio image pixels at the X-ray source optical position, using the {\sc{maxfit}} tool of CASA. We found 49 sources with radio peak flux densities $\geq 3\sigma$ (where $\sigma$ is the local noise), leading to a total of 153 radio detected X-ray sources. For the remaining 90 sources, we estimated a 3$\sigma$ radio upper limit. A summary of the X-ray/radio match analysis is reported in Table \ref{tab:cross}, where we highlight the number of radio detected and undetected sources with spectral redshift and classification. The redshift distribution of the 243 X-ray sources is shown in Fig.~\ref{fig:red_distr} (empty black histogram), with highlighted the sources with spectroscopic redshift (filled red histogram) and those, among the latter, with radio detection (filled blue histogram). We notice that the X-ray sample span a very large redshift range, with the bulk of the sources at $0<z<3$. In addition Fig.~\ref{fig:red_distr} clearly shows that the redshift range sampled by the spectroscopic sample is similar  to the one spanned by the full sample.

\begin{table}
\caption{\label{tab:cross} Summary of X-ray/radio match analysis .
}
\centering
\begin{tabular}{c|c|c|c}
\hline \hline
Redshift  & Radio & Radio & All \\
type & detections & upper limits  &  \\
 (1) & (2) & (3) & (4) \\
\hline
photometry only & 66 & 54 & 120 \\
\hline
 spectroscopy  & 87 & 36 & 123 \\

\hline
any  & 153 & 90 & 243 \\
\end{tabular}
\tablefoot{ (1) type of redshift available: only photometric or also spectroscopic. (2) number of X-ray sources detected at $\geq 3\sigma$ in the radio for a given redshift type. (3) number of X-ray sources undetected in the radio, for which a $3\sigma$ upper limit was derived, for a given redshift type. (4) total number of X-ray sources for a given redshift type.}

\end{table}

\begin{figure*}[t]
        \centering
\resizebox{\hsize}{!}
{\includegraphics[]{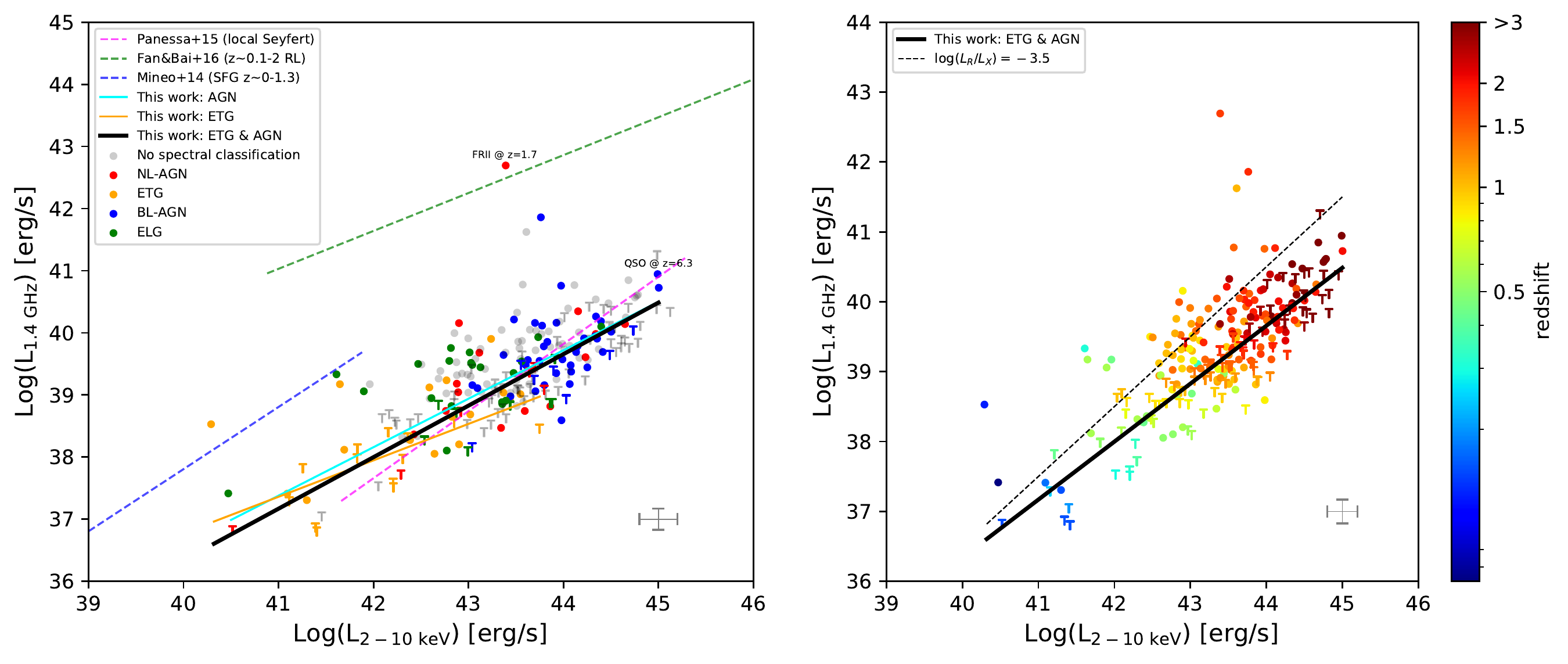}}

        \caption{X-ray intrinsic luminosity in the 2 - 10 keV band versus rest frame 1.4 GHz radio luminosity for our sample. \textit{Left}: Sources are color-coded by the source type. NL-AGN, BL-AGN, ETG and ELG are indicated by blue, red, orange and green symbols, respectively. The light-gray symbols mark the sources without a spectroscopic classification. Circles mare the radio detected sources, while $\top$ mark the 3$\sigma$ upper limits for the undetected ones. Notable sources, that are the $z=6.3$ QSO and the $z=1.7$ FRII galaxy, are labeled in the figure. The solid orange, cyan and black lines are the correlations found in this work for ETG, AGN and combined ETG/AGN samples, respectively. The dashed magenta line is the \cite{panessa_2015} correlation found for local Seyfert galaxies (slope ${\sim}$ 1.1), while the blue dashed line is the relation for $0 \lesssim z \lesssim 1.3$ SFGs \citep{mineo_2014}. The dashed green line is the relation found for $z\sim 0.1 - 2$ RL AGN by \cite{fanbai_2016}. All the relations are plotted in the X-ray luminosity range in which they have been derived. Typical uncertainties on the radio (X-ray) luminosity is indicated by the vertical (horizontal) error bar in the bottom-right corner. \textit{Right}: Same as \textit{Left} panel, except that sources are color coded by their redshift (right color-bar). Here we only show the correlation we derived for the combined ETG/AGN sample (black solid line). The black dashed line marks the $\log(L_R/L_X) =-3.5$ relation, set as the threshold between RL and RQ AGN (see text).}
        
        \label{fig:radioX}
\end{figure*}

We converted the observed flux density at $S_{1.34~\mathrm{GHz, obs}}$ to the rest-frame $\nu_\mathrm{rest}$ = 1.4 GHz luminosity as follows:
\begin{equation}\label{eq:fluxtolum}
L_{1.4~\mathrm{GHz}} = 4 \pi~D_L^2
\times \frac{S_{1.34~\mathrm{GHz, obs}}}{\nu_\mathrm{obs}^{\alpha}}
\times \left(\frac{\nu_\mathrm{rest}}{1+z}\right)^{1+\alpha}~~\mathrm{erg~s^{-1}}
\end{equation}
where $D_L$ is the luminosity distance and $\alpha$ is the spectral index, assumed equal to $-0.7$. All the quantities are expressed in cgs units. In Fig. \ref{fig:radioX} we show the position of our sources in the radio–X-ray luminosity plane. 
The X-ray luminosities were derived by \cite{marchesi_2021}: they represent intrinsic, de-absorbed luminosities measured in the 2-7 keV rest-frame and extrapolated up to 10 keV assuming a photon index $\Gamma = 1.8$ \cite[i.e., the average value for AGN, see][]{piconcelli_2005}. 
Sources are color-coded based on source type (left panel) and on source redshift (right panel). Sources with no spectral classification are shown in light-gray in the left panel. Radio detected sources are indicated by the circles, while radio upper limits are indicated by the $\top$. In the left panel we also report different X-ray/radio correlations from the literature, which refer to different types of objects. The magenta line represents the X-ray/radio relation found for a sample of local ($z\lesssim$ 0.35) X-ray-selected Seyfert galaxies by \citeauthor{panessa_2015} (\citeyear{panessa_2015}; slope $\sim$ 1.1). The blue line indicates the relation found by \cite{mineo_2014} for a sample of $z \sim 0-1.3$ SFGs, where their X-ray luminosities $L_{0.5-8~\mathrm{keV}}$ have been rescaled to the 2-10 keV band (assuming a factor $L_{2-10~\mathrm{keV}}/L_{0.5-8~\mathrm{keV}} = 1.28$, see their Sec. 4.2). The green line shows the relation found for $z\sim 0.1 - 2$ RL AGN by \cite{fanbai_2016}. We note that the FRII radio galaxy at the center of the protocluster at $z=1.7$ is located in proximity of the RL AGN relation, as expected.  
We also note that both the radio and X-ray luminosities increase with redshift, as expected for flux-limited samples (see right panel). The black dashed line in the right panel marks the radio loudness $R_X = \log(L_R/L_X) =-3.5$ relation. We follow \citealt{lambrides_2020}, who set this value as a threshold between RL and RQ AGN, on the basis of low-luminosity AGN samples (see also \citealt{terashima_2003}). We notice that most of the 243 X-ray sources are located below this threshold: specifically 82\% of the radio detected ones and 97\% of the radio upper limits. We further discuss the radio loudness distribution of the spectroscopically classified sources at the end of this section.

\begin{table}
\caption{\label{tab:corr} Summary of the X-ray/radio luminosity correlation analysis.
}
\centering
\begin{tabular}{cccccc}
\hline \hline
Type & $N$ & $\rho$ & $P_\mathrm{nh}$ & $a$ & $b$ \\
(1) & (2) & (3) & (4) & (5) & (6) \\
\hline

ETG         & 32 & 0.525 & $10^{-3}$ & 0.58 $\pm$ 0.15 & 13.31 \\
AGN         & 63 & 0.501 & $10^{-4}$ & 0.78 $\pm$ 0.19 & 5.38 \\
ETG\&AGN    & 95 & 0.688 & ${<}10^{-4}$ & 0.83 $\pm$ 0.1 & 3.17 \\
ELG         & 28 & 0.155 & $0.42$ & -- & -- \\

\hline    	  
\end{tabular}
\tablefoot{(1) Source class. (2) Number of objects in the class. (3) Spearman rank correlation coefficient. (4), null-hypothesis, that is the probability that a correlation is not present. (5) and (6) slope and coefficient of the linear regression method ($L_R = a~L_X +b$) applied to AGN-driven subsamples (see text).}

\end{table}

From the analysis of Fig. \ref{fig:radioX} (left panel), it is clear that different classes of objects tend to occupy different regions of the X-ray/radio luminosity plane: NL-AGN and ELG are preferentially clustered around intermediate values of $L_X$, in the range $42.5 \lesssim \log(L_X) \lesssim 43.5$; while ETG and BL-AGN are typically located at lower and higher X-ray luminosities, respectively. It is also interesting to note that ETG and AGN seem to show a tighter correlation between the radio and X-ray luminosity than ELG, which show a flatter radio luminosity distribution.  We explored this further, by performing a survival analysis of the various subpopulations using the {\sc{asurv}} tool \citep{feigelson_1985,isobe_1986,lavalley_1992}, that accounts also for the presence of upper limits, and calculated the Spearman rank correlation coefficient $\rho$ \citep{spearmann_1904} and the corresponding probability that a correlation is not present (null hypothesis, $P_\mathrm{nh}$). We recall that the Spearman coefficient increases for positive correlations (up to $+1$ for a perfect monotone increasing function), while a value of 0 indicates no correlation. This analysis could obviously be carried out only for the 123 sources  for which \cite{marchesi_2021} provided the spectral classification. We investigated the significance of the correlation between X-ray and radio luminosities for the following subsamples: ETG (32 objects), AGN (i.e., BL-AGN and NL-AGN, 63 objects) and ELG (28 objects).

The results are summarized in Table~\ref{tab:corr} (columns 3 and 4). We obtained $\rho$ = 0.525 and $\rho$ = 0.502 for ETG and AGN, respectively. The null hypothesis probability in both cases is very low ($P_\mathrm{nh} = 10^{-3}$ and $P_\mathrm{nh} = 10^{-4}$ for ETG and AGN, respectively), implying a significant X-ray/radio luminosity correlation for these classes of sources. 
These findings suggest that a common mechanism is at the origin of the emission observed in the two bands for ETG and AGN (likely associated with SMBH accretion). If we combine the ETG and AGN samples we obtain an even stronger correlation ($\rho$ = 0.688) with respect to those of the two separated samples. This strengthening is likely the result  of better statistics and a larger range of X-ray luminosities spanned by the combined sample. We caveat that luminosity-luminosity correlations may reflect luminosity-distance correlations in flux-limited samples (see right panel in Fig. \ref{fig:radioX}). To check that this is not the case, we performed a partial Kendall $\tau$ correlation test \citep{akritas_1996,merloni_2006}, which provides partial correlation coefficient and significance by taking into account three variables (i.e., $L_R$, $L_X$ and $z$). We found that the correlation cofficient between $L_R$ and $L_X$, once accounting for their correlation with $z$, is $\tau=0.32$, corresponding to $P_\mathrm{nh} \sim 2.5 \times 10^{-6}$. This  rules out that the observed $L_R$/$L_X$ correlation is driven by distance. 
As far as ELG are concerned, we obtain $\rho$ = 0.155 and $P_\mathrm{nh} = 0.42$, resulting in a high probability of no correlation between radio and X-ray emission. This suggests that ELG include a significant fraction of sources where radio and X-ray luminosities are to be ascribed to different emission mechanisms. We return on this point later on. 

To quantify the correlation for the AGN-driven subsamples, we applied the regression method by \cite{buckley_1979} using {\sc{asurv}}, taking in account the upper limits. The correlation is described by a linear function of the form:
\begin{equation}\label{eq:RX_rel}
\log(L_R)= a ~\log(L_X) + b
\end{equation}
where $L_R$ and $L_X$ are expressed in erg/s. Results are reported in columns 5 and 6 of Table \ref{tab:corr}. We note that the slope of the ETG correlation ($a=0.58 \pm 0.15$) is well in the range of the slopes found for low luminosity AGN in the fundamental plane  (${\sim}0.5-0.7$, \citeauthor{merloni_2003} \citeyear{merloni_2003}, see also \citeauthor{dong_2021} \citeyear{dong_2021}). Such slope values are commonly interpreted as a signature of radiatively inefficient accretion processes \cite[see, e.g.,][see also Sec. \ref{sec:intro}]{narayan_1994,dong_2014}. This interpretation is also consistent with an ETG spectral classification  (i.e., with the fact that AGN spectral features are not observed), as well as with the fact that many of them have low X-ray luminosities ($<10^{42}$ erg s$^{-1}$). Our ETG sample, which mostly comprises nearby sources ($z<0.5$; see Fig.~\ref{fig:radioX}), can be directly compared with the nearby sample of \citeauthor{panessa_2015} (\citeyear{panessa_2015}; magenta line in Fig. \ref{fig:radioX}). The steep slope found for the latter ($a=1.08\pm 0.15$) is generally interpreted as a signature of efficient accretion processes \cite[][see also Sec. \ref{sec:intro}]{coriat_2011,dong_2014}. Indeed these objects are mainly (type 1 and type 2) Seyfert galaxies, and generally feature higher X-ray luminosities ($>10^{42}$ erg s$^{-1}$, up to $10^{45}$ erg s$^{-1}$) than those of our ETG sample. Despite the large uncertainties of the measured slopes (${\sim} 0.1-0.15$), we can conclude that our ETG sample likely probes  a different accretion efficiency regime with respect to \cite{panessa_2015} sample. As for the AGN subsample, we measure a slope $a=0.78 \pm 0.19$. Given its large uncertainty, this value can be considered consistent with both the  ETG and the \cite{panessa_2015} results. Nevertheless, a steeper slope with respect to the ETG sample is expected, as more efficient accretion processes should occur in these objects. If confirmed, a possible explanation for an intermediate slope between low and high radiatively efficient AGN could be that our AGN sample consists of a mix of objects with different accretion efficiency, or that a large fraction of the sources in our AGN sample have intermediate efficiency (similar to that exhibited by some X-ray binaries, \citealt{gallo_2012}). A redshift evolution scenario, implying a flattening of the slope with redshift, could also be considered, as our AGNs are mostly at $z>1$ (see Fig.~\ref{fig:radioX}). However, the hypothesis of mixed populations is more likely, and is supported by the fact that the AGN+ETG sample displays a similar slope ($0.83\pm 0.1$).

\begin{figure}[t]
        \centering
\resizebox{\hsize}{!}
{\includegraphics[]{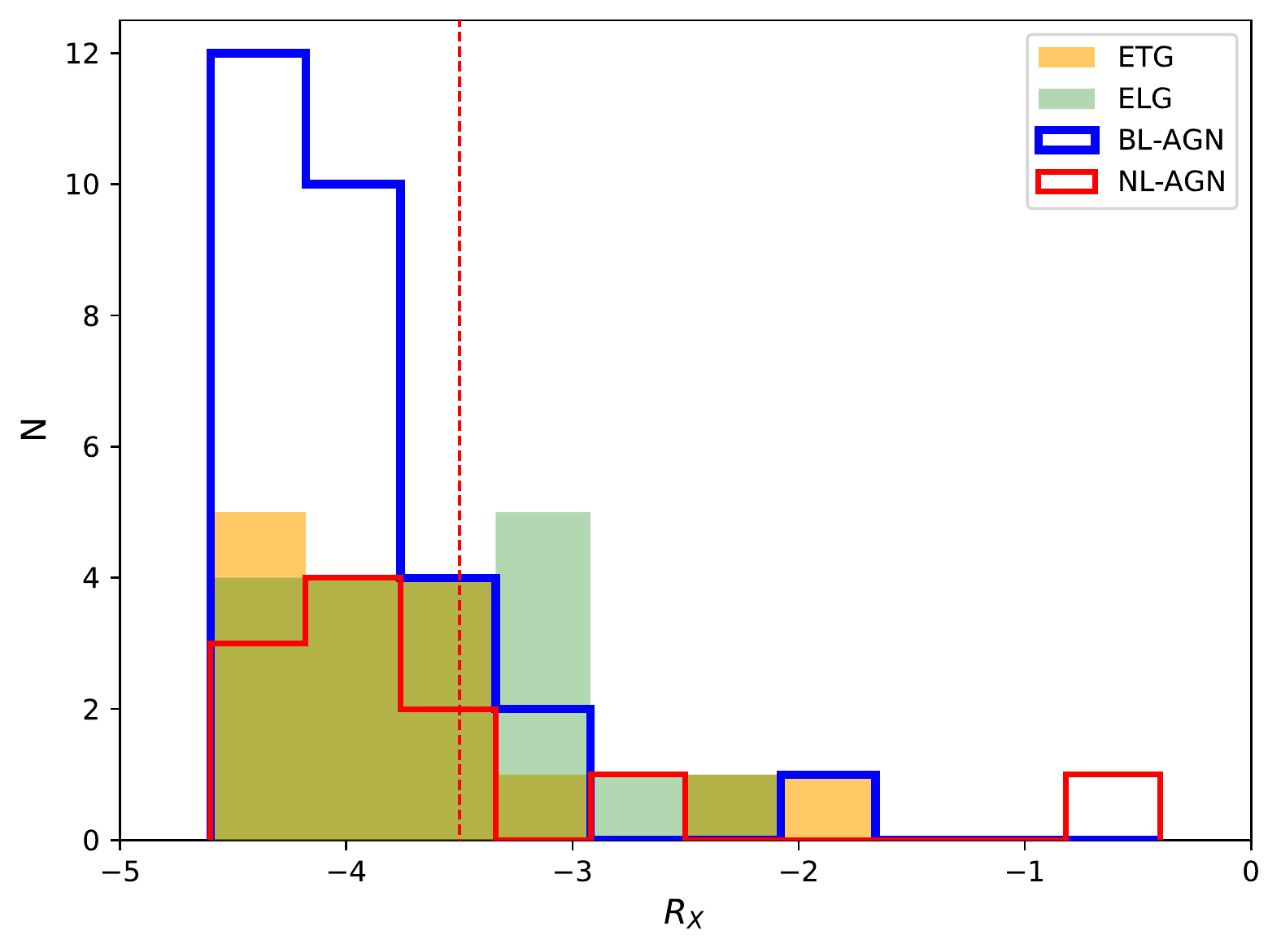}}

        \caption{Radio loudness distribution (in bins of 0.4) of the radio detected sources of the X-ray spectroscopic sample, color-coded by their spectral classification. The $R_X$ = 3.5 threshold used to separate RQ from RL AGN \citep{lambrides_2020} is indicated by the red dashed vertical line. The most RL NL-AGN is the FRII galaxy at $z=1.7$.}
        
        \label{fig:RX}
\end{figure}

\begin{figure}[t]
        \centering
\resizebox{\hsize}{!}
{\includegraphics[]{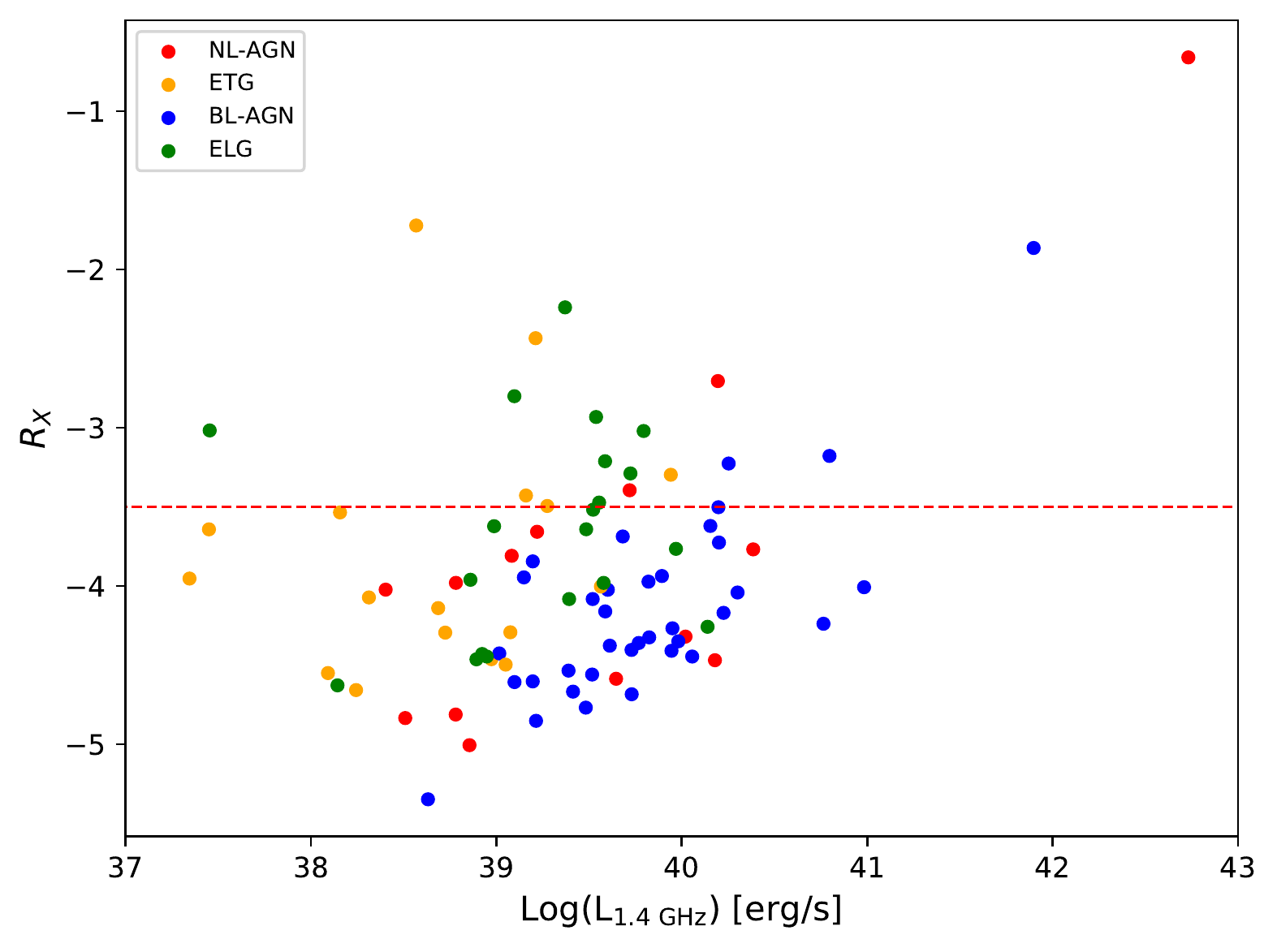}}

        \caption{Radio loudness of the radio detected sources of the X-ray spectroscopic sample as a function of the 1.4 GHz radio luminosity, color-coded by the objects class.}
        
        \label{fig:RX_Lrad}
\end{figure}

In the following we analyse the radio loudness $R_X = \log(L_R/L_X)$ of the sources in our sample with a spectroscopic classification. Fig. \ref{fig:RX} shows the $R_X$ distribution for each class of objects (we only show here radio detected sources). It is clear that most of the AGN-driven (ETG+AGN) sources in our sample are radio quiet, meaning that they have $R_X \lesssim -3.5$ (${\sim}$83\%, or $\gtrsim$87\% when we include radio upper limits). 
The most RL AGN corresponds to the FRII galaxy at the center of the protocluster at $z=1.7$. In addition, a significant fraction of the radio detected ELG (${\sim}$40\%) display $R_X \gtrsim -3.5$, meaning that they apparently host RL AGN. 
In Fig. \ref{fig:RX_Lrad} we show $R_X$ as a function of $L_R$ for the radio detected sources; the ELG which result radio loud show a radio luminosity comparable with, or lower than, the one of RQ AGN. We again argue that these ELG are not genuinely radio loud, but rather display  radio and X-ray emission of different origin. As a matter of fact, most ELG lie between the RQ AGN and the SFG radio/X-ray relations in Fig.~\ref{fig:radioX}, and very far from the RL AGN one, meaning that none of them can describe their radio/X-ray properties. Indeed, the ELG in our sample show high X-ray luminosities  ($> 10^{42}$ erg/s), typical of AGN, and radio powers that can reach L$_R \sim 10^{39-40}$ erg/s.
We calculated the SFR for each source by assuming their $z_{spec}$ and radio luminosity, and obtained a SFR range ${\sim}20 -180~\mathrm{M_{\odot}}$/yr. This is consistent with the main-sequence SFR expected in the redshift range $1\lesssim z \lesssim 2$ \citep{schreiber_2015} where most of these sources lie, strengthening the hypothesis of SF being responsible of the observed radio emission.

\section{Summary and conclusions}
\label{sec:concl}
We presented deep L-band (1.34 GHz effective frequency) JVLA observations of the J1030 equatorial field. The reached sensitivity (median rms noise $\lesssim$ 3.4 $\mu$Jy/b) makes this one of the deepest extragalactic radio survey to date. Using the PyBDSF tool, we extracted a catalog of 1489 radio sources with S/N $>$ 5 and performed reliability analysis, finding that the detections are reliable at $\geq$ 93\%. We derived the source counts as a function of the flux density down to $\sim$ 20 $\mu$Jy, finding overall consistency with recent determinations and models. The counts show a slight excess at flux densities $\sim 50$ $\mu$Jy, possibly associated with the presence of known overdensities in the field. 

Thanks to the depth of our observations,  we detected for the first time in the radio band the $z=6.3$ QSO SDSS J1030+0524 at the center of the field, measuring a flux density of $S_{1.34~\mathrm{GHz,~obs}} = 26 \pm 5 ~\mu$Jy, corresponding to a 1.4 GHz rest-frame luminosity $L_{1.4~\mathrm{GHz,~rest}} = (6.2 \pm 1.2) \times 10^{24}$ W/Hz, assuming $\alpha=-0.7$. From the radio and the UV rest-frame luminosity ($L_{4400~\mathrm{\AA,~rest}}$ ${\sim}$ 12.91 $L_{\odot}$) reported by \cite{banados_2015}, we derived an optical radio loudness $R_O =$ 0.62 $\pm$ 0.12, which classifies our source as the most RQ AGN and the faintest radio detected QSO at $z\gtrsim6$ discovered to date. Remarkably, this object is at the center of the most distant spectroscopically confirmed overdensity known to date \citep{mignoli_2020}.

The depth of our survey also allowed us to unveil the presence of extended diffuse radio emission in the lobes of the FRII radio galaxy at the center of the $z=1.7$ protocluster in the J1030 field. Based on this, we revised the source size to 700 kpc, making it a giant radio galaxy. 
The FRII galaxy  total flux density has been measured to be $S_{1.34~\mathrm{GHz,~obs}} = 25 \pm 4 ~\mu$Jy,  corresponding  to a 1.4 GHz rest-frame luminosity $L_{1,4~\mathrm{GHz,~rest}} = (3.5 \pm 0.7) \times 10^{26}$ W/Hz (assuming again $\alpha=-0.7$). 
On the basis of the jet and counter-jet length and flux ratios, we estimated an inclination angle with respect to the line of sight of $\theta{\sim}80^\circ$, which is consistent with the one derived by \cite{gilli_2019}, on the basis of previous shallower VLA observations (\citealt{petric_2003}). The newly discovered extended emission is nicely cospatial with spots of X-ray diffuse emission detected around the FRII galaxy lobes, indicating a possible interaction between the radio galaxy and the surrounding ICM. 

Furthermore, we detected for the first time in the radio band two additional gas-rich members of the protocluster. For one of them we could perform SED fitting, from which we derived a SFR  $<$43 $\mathrm{M_{\odot}}$/yr. This upper limit is consistent with the range derived by \cite{damato_2020b} from an ALMA-based molecular gas mass measurement and exploiting the SK law (${\sim}40-60~\mathrm{M_\odot}$/yr). It is also consistent with the typical SFR of main-sequence galaxies with similar stellar mass ($\log(M_{\ast}/\mathrm{M_{\odot}}) = 10.75 \pm 0.25$) and redshift  \citep{schreiber_2015,genzel_2015}. On the other hand, this SFR is at least a factor $2\times$ too low to explain the observed JVLA flux density, suggesting that at least part of the observed radio emission is of AGN origin. If confirmed, this would be the second radio AGN detected in the protocluster. 

Finally, we exploited the deep (${\sim}$500 ks) \textit{Chandra} observations of the J1030 field to explore the radio/X-ray luminosity correlation of RQ AGN. To this end we matched the 243 X-ray sources with photometric and/or spectroscopic redshifts \citep{marchesi_2021} with our radio catalog, and pushed the search for radio counterparts down to $3\sigma$ for sources with no association in the radio catalog. As a result we found 153 radio detections. For the remaining sources we estimated $3\sigma$ radio upper limits. 
The sources span a redshift range  $0 \lesssim z \lesssim 3$, with a high redshift tail extending up to $z=6.3$ (the redshift of the SDSS J1030+0524 QSO). We calculated the rest-frame 1.4 GHz luminosity of all the 243 X-ray sources, 
and investigated the X-ray/radio luminosity correlation for the 123 sources (87 radio detected and 36 with radio upper limits) provided with spectroscopic redshift and classification, through survival analysis. We found that AGN (i.e., BL-AGN and NL-AGN) and ETG feature a significant X-ray/radio correlation (null-hypothesis probability $P_\mathrm{nh}\leq 10^{-3}$), indicating that a common mechanism is at the origin of the emission observed in the
two bands (likely associated with SMBH accretion). 
In addition, most ETG and AGN (${\sim}$83\%, or ${\gtrsim}$87\% if we include radio upper limits) show a radio-to-X-ray radio loudness $R_X \lesssim -3.5$, classifying these objects as RQ AGN \citep{terashima_2003,lambrides_2020}. 
We found that ETG (most of which have $L_X<10^{42}$ erg s$^{-1}$ and $z<0.5$) follow a relation $\log(L_R)= a ~\log(L_X) + b$ with a slope $a\sim 0.6$, in agreement with fundamental plane studies of low luminosity AGN 
(slope $0.5 − 0.7$, \citealt{merloni_2003,dong_2021}). This slope is commonly interpreted as a signature of inefficient accretion processes, in contrast with what is found for luminous ($L_X\gtrsim10^{43}$ erg s$^{-1}$) X-ray-selected local Seyfert galaxies, for which \cite{panessa_2015} found a steeper relation ($a{\sim}$1.1) that is interpreted as a signature of efficiently accreting systems. Despite the large uncertainties of the measured slopes (${\sim}$0.1--0.15), we can conclude that our ETG sample likely probes a different accretion efficiency regime with respect to \cite{panessa_2015} sample. 
The AGN subsample displays an intermediate correlation ($a\sim 0.8$) between local ETG and Seyfert galaxies. The 1$\sigma$ uncertainty of the derived slope is large (${\sim}0.2$), but, if confirmed, this result may suggest that our AGN sample consists of a mix of objects with different accretion efficiency and/or that a large fraction of them have intermediate efficiency. 
Finally, we investigated the properties of ELG, finding that these sources do not show a significant X-ray/radio correlation ($P_\mathrm{nh}=0.42$). Our analysis suggests that star formation and SMBH accretion are responsible for the radio and X-ray emission in these objects, respectively. 

\begin{acknowledgements}
QD acknowledges support from European Social Fund, Project FP2012369501/2020: ``Assegni di ricerca FSE SISSA 2020''. IP and QD acknowledge support from INAF under the SKA/CTA PRIN ``FORECaST'' and the PRIN MAIN STREAM ``SAuROS'' projects. The National Radio Astronomy Observatory is a facility of the National Science Foundation operated under cooperative agreement by Associated Universities, Inc.
\end{acknowledgements}



%
   \bibliographystyle{aa} 
  \bibliography{jvla} 
%

\end{document}